\documentclass[twocolumn,showpacs,amsmath,amssymb,floatfix,a4paper]{revtex4}
\usepackage{xspace}
\usepackage{txfonts}
\usepackage{color}
\usepackage{graphicx}
\usepackage{dcolumn}
\newcolumntype{Z}[1]{D{.}{\cdot}{#1} }
\usepackage{bm}
\usepackage{longtable}

\usepackage{array}
\newcommand{\dr}{{\mathrm d}^{3}{\mathbf r}}

\newcommand{\rr}{{\mathbf r}}

\newcommand{\etal}{\emph{et al.}\xspace}

\newcommand{\oQ}[2]{\ensuremath{\hat{Q}^{#1}_{#2}}\xspace}

\newcommand{\T}[2]{\ensuremath{T^{#1}_{#2}}\xspace}
\newcommand{\A}[2]{\ensuremath{\alpha^{#1}_{#2}}\xspace}

\newcommand{\Au}[2]{\ensuremath{\alpha^{#1}_{#2}(i u)}\xspace}
\newcommand{\IIu}[2]{\ensuremath{I^{#1}_{#2}(i u)}\xspace}
\newcommand{\JJu}[2]{\ensuremath{J^{#1}_{#2}(i u)}\xspace}

\newcommand{\gra}[0]{\ensuremath{\alpha}\xspace}
\newcommand{\grb}[0]{\ensuremath{\beta}\xspace}
\newcommand{\grg}[0]{\ensuremath{\gamma}\xspace}
\newcommand{\grd}[0]{\ensuremath{\delta}\xspace}
\newcommand{\gre}[0]{\ensuremath{\epsilon}\xspace}
\newcommand{\grp}[0]{\ensuremath{\phi}\xspace}

\newcommand{\Edisp}[1]{\ensuremath{E^{(#1)}_{\rm disp}}\xspace}
\newcommand{\EdispMP}[1]{\ensuremath{E^{(#1)}_{\rm disp,MP}}\xspace}

\newcommand{\U}[1]{\ensuremath{u_{#1}\xspace}}
\newcommand{\fPIe}{\ensuremath{4\pi \epsilon_{0}}}

\newcommand{\CamCASP}{{\sc CamCASP}\xspace}

\newcommand{\abinitio}{{\em ab initio}\xspace}

\newcommand{\distortion}{\ensuremath{\eta}\xspace}

\newcommand{\HH}[1]{(H$_2$)$_{#1}$\xspace}

\mathchardef\lt="313C \mathchardef\gt="313E
\mathcode`<="4268 \mathcode`>="5269
\begin{document}

\title{Anomalous non-additive dispersion interactions in systems of three 
       one-dimensional wires}
\author{Alston J.\ Misquitta}
\affiliation{School of Physics and Astronomy, Queen Mary, University of London,
London E1 4NS, UK}
\author{Ryo Maezono}
\affiliation{School of Information Science, JAIST, Asahidai 1-1, Nomi,
  Ishikawa 923-1292, Japan.}
\author{Neil D.\ Drummond}
\affiliation{Department of Physics, Lancaster University, Lancaster LA1 4YB, UK}
\author{Anthony J.\ Stone}
\affiliation{The University Chemical Laboratory, Lensfield Road, 
Cambridge CB2 1EW, UK}
\author{Richard J.\ Needs}
\affiliation{TCM Group, Cavendish Laboratory, 19, J.\ J.\ Thomson Avenue,
Cambridge CB3 0HE, UK }

\date{\today}


\begin{abstract}
The non-additive dispersion contribution to the binding energy of
three one-dimensional (1D) wires is investigated using wires modelled
by (i) chains of hydrogen atoms and (ii) homogeneous electron gases.
We demonstrate that the non-additive dispersion contribution to the
binding energy is significantly enhanced compared with that expected
from Axilrod-Teller-Muto-type triple-dipole summations and follows a
different power-law decay with separation.
The triwire non-additive dispersion for 1D electron gases scales according to
the power law $d^{-\beta}$, where $d$ is the wire separation, with
exponents $\beta(r_s)$ smaller than 3 and slightly increasing with
$r_s$ from 2.4 at $r_s = 1$ to 2.9 at $r_s=10$, where $r_s$ is the
density parameter of the 1D electron gas.
This is in good agreement with the exponent $\beta=3$ suggested by the
leading-order charge-flow contribution to the triwire non-additivity,
and is a significantly slower decay than the $\sim d^{-7}$ behaviour
that would be expected from triple-dipole summations.
\end{abstract}

\maketitle

\section{Introduction}
\label{sec:introduction}

 
Recently there has been a resurgence in attempts to model the dispersion
interaction between low-dimensional nano-scale objects more accurately.
Using an array of electronic structure \cite{Spencer09:thesis,MisquittaSSA10,
DrummondN07} and analytical \cite{DobsonWR06} techniques, several groups have 
demonstrated that the dispersion interaction between one- and two-dimensional
systems can deviate strongly from that expected from the well-known
{\em additive} picture of $r^{-6}$-type interactions 
\cite{Stone:book:13,Kaplan05:book}.
For the case of parallel one-dimensional (1D) metallic wires separated by distance $d$,
Dobson \etal \cite{DobsonWR06} demonstrated that the van der Waals dispersion
interaction should decay as $\sim - d^{-2}[\ln (\gamma d)]^{-3/2}$, where
$\gamma$ is a constant that depends on the wire width.
This analytic result was subsequently verified by Drummond and Needs
\cite{DrummondN07} using diffusion quantum Monte Carlo (DMC) calculations
\cite{FoulkesMNR01}.  This change in the power-law of the dispersion energy 
can be understood as arising from correlations in extended plasmon modes 
in the metallic wires \cite{DobsonWR06,Dobson07a,DobsonMcLRWGLD01}.
These plasmon modes would be expected in any low-dimensional system with a
delocalized electron density.

Misquitta \etal \cite{MisquittaSSA10} have recently extended these results
to the more general case of finite- and infinite-length wires with arbitrary
band gap. Using dispersion models that include non-local charge-flow 
polarizabilities they were able to describe the dispersion interactions in all
cases, including the insulating and semi-metallic wires. In these models the plasmon-like
fluctuations are modelled by the charge-flow polarizabilities which, at 
lowest order, result in a $-d^{-2}$ dispersion interaction 
\cite{Stone:book:13,MisquittaSSA10}. For metallic wires these terms are dominant
at all separations and yield the result of Dobson \etal for the dispersion.

Curiously, many of these results were known as early as 1952. Using a tight-binding
H\"{u}ckel-type model for linear polyenes, Coulson and Davies \cite{CoulsonD52}
investigated the dispersion interactions between the chains in a variety of 
configurations and with a range of highest occupied to lowest unoccupied molecular
orbital (HOMO--LUMO) gaps. Their conclusions about the non-additivity of the
dispersion interaction and the changes in power law (deviations from the expected
effective $-d^{-5}$ London behaviour) are essentially
identical to those reached by Misquitta \etal \cite{MisquittaSSA10}. A few years
later Longuet-Higgins and Salem \cite{Longuet-HigginsS61} reached similar conclusions
and related the non-additivity of the dispersion to the existence of long-range
correlations within the system. A decade later Chang \etal \cite{ChangCDY71} 
used Lifshitz theory to derive an analytic form of the dispersion interaction
between two metallic wires that is identical to the expression of 
Dobson \etal \cite{DobsonWR06}, though the latter considered many more cases. 

The current interest in this field stems from two sources.
First we have recently witnessed an explosion of work on nano-scale devices 
confined in one or two dimensions. Examples are carbon nanotubes and devices 
based on graphene and related materials. To model accurately the self-assembly
of these materials we need to describe correctly their interactions, particularly
the ubi\-qui\-tous dispersion interaction.
Second, \abinitio electronic structure methods have now achieved a level of
accuracy and computational efficiency that allows them to be applied to such systems. 
These methods have exposed the inadequacies of assumptions and approximations
made in many empirical models. From the research cited above we now know that 
the dispersion energy exhibits much more substantial non-additivity than assumed 
previously. 

We emphasise here that empirical models for the dispersion energy prove inadequate
because they rely on the assumption of additivity through the pair-wise
$C_6^{ab}/r_{ab}^6$ model with van der Waals coefficients $C_6^{ab}$ between
sites $a$ and $b$ assumed to be isotropic constants, with little or no variation
with changes in chemical environment. 
Part of the missing non-additivity arises from the local
chemical environment changes and from through-space coupling between the dipole
oscillators. The remainder arises from the metallic-like contributions that
are responsible for the anomalous dispersion effects that are the subject of
this paper. We stress that while the first kind of non-additivity can be described
by coupled-oscillator models \cite{GobreT13} and \abinitio derived dispersion
models such as those obtained from the Williams--Stone--Misquitta 
\cite{MisquittaS08a,MisquittaSP08} effective local polarizability models,
as we shall see next, the latter, that is, the non-additivity arising from 
metallic contributions, requires models that take explicit account of 
extended charge fluctuations.

\begin{figure}
  \includegraphics[width=0.4\textwidth,clip]{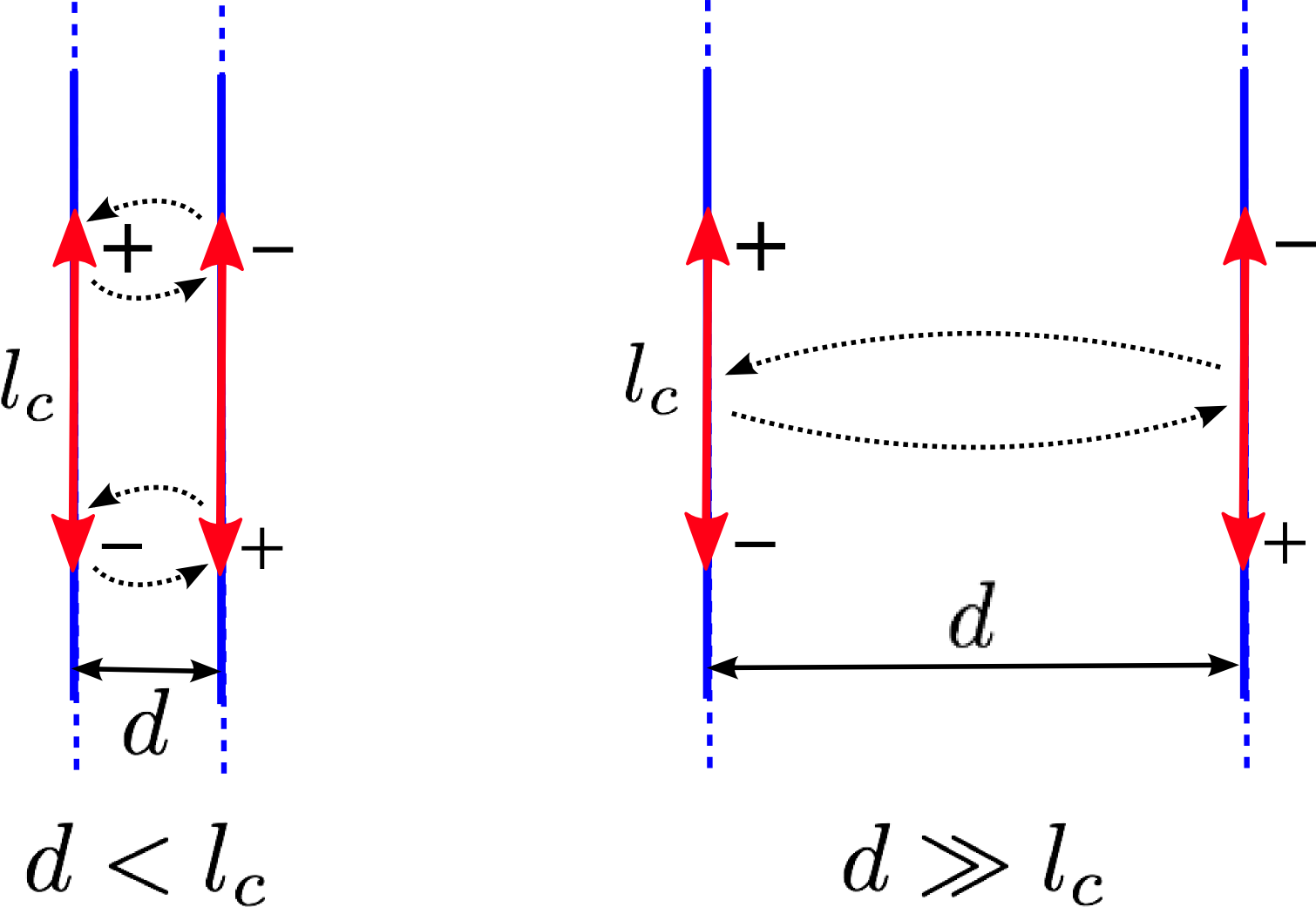}
  \caption[Physical origin of anomalous dispersion energy between two parallel
  1-D wires]{
  Electronic fluctuations in (infinite) 1D wires (in blue) arise from
  the tightly bound electrons (not shown) and electrons at the band edge 
  (represented by the red arrows). The extent of these fluctuations will 
  depend on the band gap (see text) and will have a typical length scale
  $l_c$. 
  An extended fluctuation of $+\cdots-$ in one wire will induce
  a $-\cdots+$ fluctuation in the other.
  If $d$ is the separation, we can identify two cases: (1) $d \lt l_c$ and
  (2) $d \gg l_c$. As explained in the text, the leading-order dispersion
  interaction in the former is associated with charge-induced-charge
  interactions, and that of the latter with dipole-induced-dipole interactions. 
  \label{fig:edisp2-physics}
  }
\end{figure}

The unusual nature of the second-order dispersion
energy, \Edisp{2}, for infinite, parallel 1D wires of arbitrary
band gap can be understood as follows. The electronic fluctuations in the 
wire are broadly of two types: the short-range fluctuations associated 
with tightly bound electrons and the long-range plasmon-type fluctuations
associated with electrons at the band edge. The former give rise to the
standard dispersion model while the latter are responsible for the effects
discussed in this paper and those cited above. For systems with a finite
gap, the plasmon-like modes will be associated with a finite length scale,
$l_c$, defined, for example, via the Resta localization tensor \cite{Angyan09a}.
For metallic systems this length scale is expected to diverge. 
Consider now the two cases depicted in Fig.~\ref{fig:edisp2-physics}.
In the first case the wires are separated by $d \lt l_c$. Here, the 
leading-order contribution from the spontaneous extended fluctuation
depicted in the figure is that between charges and leads to the $-d^{-2}$
behaviour of \Edisp{2}: the spontaneous fluctuation at the first wire
results in a field $\sim d^{-1}$ at the second and this interacts with the
first via another $d^{-1}$ interaction leading to the favourable $-d^{-2}$
dispersion energy. 
Only local charge-pairs contribute to this leading order interaction,
consequently the dispersion interaction per unit length remains $-d^{-2}$.

If, on the other hand, $d \gg l_c$, the extended fluctuation
at the first wire generates a dipole field of strength $\sim d^{-3}$ at the second,
and the resulting induced (extended) dipole interacts with the first via 
a dipole-dipole interaction leading to another factor of $d^{-3}$.
This gives a nett favourable dispersion interaction of $-d^{-6}$.
In this case, to find the nett dispersion interaction per unit wire length
we need to sum over all the interactions between an element of one
wire and all elements of the other, which
leads to an effective $-d^{-5}$ dispersion interaction just as for the
point-like fluctuating dipoles of the tightly-bound electrons
\cite[p.173]{Parsegian05:book}.
In both cases, the usual $-d^{-5}$ effective dispersion interaction
from the tightly bound electrons must be included too.

The length-scale $l_c$ is expected to diverge in a metal, leading to a single power law
$-d^{-2}$ for \Edisp{2}. For finite-gap wires we expect the two regimes
described above. This is exactly the conclusion reached by Misquitta \etal
\cite{MisquittaSSA10} and, much earlier, by Coulson and Davies \cite{CoulsonD52}.

The second-order dispersion energy is, however, only part of the story. For a group 
of interacting monomers (possibly of different types) the dispersion energy
includes contributions from second-order as well as third- and higher-order terms.
The third-order dispersion includes two- and three-body terms \cite{Stogryn71a};
the former will be denoted by $\Edisp{3}[2]$ and the latter by $\Edisp{3}[3]$.
$\Edisp{3}[2]$ is expected to be important for small-gap systems,
since these are associated with large hyperpolarizabilities, but we may
expect \emph{a priori} that as long as $\Edisp{3}[3]$ decays slowly enough
with trimer separation, it is the three-body non-additive energy
$\Edisp{3}[3]$ that will be the dominant contributor in the condensed
phase due to the far larger number of trimers compared with dimers.

The three-body non-additive energy $\Edisp{3}[3]$ is usually modelled using
the triple-dipole Axilrod--Teller--Muto expression (see Sec.\ \ref{sec:theory}) 
\cite{AxilrodT43,Muto43} from which $\Edisp{3}[3] \sim R^{-9}$, that is, the 
non-additivity decays very rapidly with separation. 
As will be demonstrated below, this expression is not valid for small-gap
systems; instead a more general expression is derived that includes contributions
from correlations between the long-wavelength plasmon-like modes. 
From the physical picture of the second-order dispersion energy given above
we may {\em a priori} expect that the true $\Edisp{3}[3]$ will be qualitatively 
different from that suggested by the triple-dipole expression. As we shall see
below, this is indeed the case.

The multipole expansion is a powerful method, but it would be reassuring to
verify its predictions using a non-expanded \textit{ab initio} approach.
In order to obtain hard numerical data describing the nonadditivity of the
dispersion interactions between metallic wires, we have evaluated the binding
energy of
three parallel, metallic wires in an equilateral-triangle configuration using
the variational and diffusion quantum Monte Carlo (VMC and DMC) methods.  VMC
allows one to take expectation values with respect to explicitly correlated
many-electron wave functions by using a Monte Carlo technique to evaluate the
multidimensional integrals.  The DMC method projects out the ground-state
component of a trial wave function by simulating drift, diffusion, and
branching processes governed by the Schr\"{o}dinger equation in imaginary
time.  In our QMC calculations each wire was modelled as a 1D homogeneous
electron gas (HEG)\@.  The dependences of the biwire and triwire interactions
on the wire separation $d$ were evaluated in order to determine the asymptotic
power law for the interaction and the non-additive three-body contribution.
We find that the long-range non-additivity is repulsive and scales as a power
law in $d$ with an exponent slightly less than three.

The paper is organised as follows. The underlying theory is described in
Sec.\ \ref{sec:theory}.  In Sec.\ \ref{sec:results} we describe the
computational details and present our results.  Finally, we discuss the
physical consequences of our results in Sec.\ \ref{sec:discussion}.

\section{Theory}
\label{sec:theory}


The non-expanded 3-body, non-additive dispersion energy has been shown to be \cite{Stogryn71a}
(all formulae will be given in SI units, but results will be in atomic units)
\begin{align}
  \Edisp{3}[3] & = -\frac{\hbar}{\pi(\fPIe)^3} \int_{0}^{\infty} du \int 
                    \dr_{1} \dr_{1'} \dr_{2} \dr_{2'} \dr_{3} \dr_{3'} \nonumber \\
        &    \frac{\alpha^{A}(\rr_{1},\rr_{1'};iu) 
                   \alpha^{B}(\rr_{2},\rr_{2'};iu)
                   \alpha^{C}(\rr_{3},\rr_{3'};iu)}
                  {|\rr_{1'}-\rr_{2}| |\rr_{2'}-\rr_{3}| |\rr_{3'}-\rr_{1}|}.
  \label{eq:Edisp33}
\end{align}
Here $\alpha^{X}(\rr_{1},\rr_{1'};iu)$ is the frequency-dependent density susceptibility
(FDDS) function for monomer $X$ evaluated at imaginary frequency $iu$
\cite{Longuet-Higgins65,Stone85}. 
The sign of the above expression has been chosen so that the polarizability tensor 
defined as
\begin{align}
  \A{aa'}{\gra\gra'}(\omega) = - \iint \oQ{a}{\gra}(\rr_1) \alpha(\rr_{1},\rr_{1'};\omega)
                                   \oQ{a'}{\gra'}(\rr_{1'}) \dr_{1}\dr_{1'}
        \label{eq:pol}
\end{align}
is positive-definite.
Here $\oQ{a}{\gra}$ is the multipole moment operator for site $a$  with component
$\gra=00,10,11c,11s,\cdots$ using the notation described by Stone 
\cite{Stone:book:13}.
As defined, $\A{aa'}{\gra\gra'}(\omega)$ is the {\em distributed} polarizability
for sites $a$ and $a'$. It describes the linear response of the expectation
value of the local operator $\oQ{a}{\gra}$ to the frequency-dependent 
(local) perturbation $\oQ{a'}{\gra'} \cos(\omega t)$ \cite{MisquittaS06}.
That is, the distributed polarizability $\A{aa'}{\gra\gra'}(\omega)$ describes
the first-order change in multipole moment of component $\gra$ at site $a$ in
response to the frequency-dependent perturbation of component $\gra'$ at a site $a'$.

For the sake of clarity we will use the following notation in subsequent 
expressions: sites associated with monomers $A$, $B$ and $C$ will be
designated by $a,a'$, $b,b'$ and $c,c'$, and angular
momentum labels by $\gra,\gra'$, $\grb,\grb'$ and $\grg,\grg'$, respectively. 
Molecular labels are hence redundant and will be used only if there is the possibility of 
confusion.

The multipole expansion of $\Edisp{3}[3]$ is obtained by expanding the
Coulomb terms in Eq.~\eqref{eq:Edisp33} as follows
\begin{align}
  \frac{1}{|\rr_1 - \rr_2|} = \oQ{a}{\gra}(\rr_1) \T{ab}{\gra\grb} \oQ{b}{\grb}(\rr_2),
  \label{eq:multipole}
\end{align}
where $\T{ab}{\gra\grb}$ is the interaction function \cite{Stone:book:13}
between multipole $\gra$ on site $a$ (in subsystem $A$) and multipole 
$\grb$ on site $b$ (in subsystem $B$).
At lowest order, the interaction function $\T{ab}{00,00} = |\rr^a-\rr^b|^{-1}$ describes
the interaction of the charge on $a$ with that on $b$.
With this multipole expansion (MP) Eq.~\eqref{eq:Edisp33} takes the form
\begin{align}
  \Edisp{3}[3] & \rightarrow \EdispMP{3}[3] = 
   + \frac{\hbar}{\pi(\fPIe)^3} 
        \T{a'b}{\gra'\grb} \T{b'c}{\grb'\grg} \T{c'a}{\grg'\gra} ~ \times \nonumber \\
   &  \int_0^{\infty} \left[ \iint \dr_1 \dr_{1'} \oQ{a}{\gra}(\rr_1) \alpha^{A}(\rr_1,\rr_{1'};iu) 
                                                  \oQ{a'}{\gra'}(\rr_{1'}) \right] \nonumber \\
   &  ~~~~~~~  \left[ \iint \dr_2 \dr_{2'} \oQ{b}{\gra}(\rr_2) \alpha^{B}(\rr_2,\rr_{2'};iu) 
                                           \oQ{b'}{\gra'}(\rr_{2'}) \right] \nonumber \\
   &  ~~~~~~~  \left[ \iint \dr_3 \dr_{3'} \oQ{c}{\grg}(\rr_3) \alpha^{C}(\rr_3,\rr_{3'};iu) 
                                           \oQ{c'}{\grg'}(\rr_{3'}) \right] du \nonumber \\
   & = + \frac{\hbar}{\pi(\fPIe)^3} 
        \T{a'b}{\gra'\grb} \T{b'c}{\grb'\grg} \T{c'a}{\grg'\gra} \nonumber \\
   & ~~~~~~~~ \times ~
        \int_0^{\infty} \A{aa'}{\gra,\gra'}(iu) \A{bb'}{\grb\grb'}(iu) 
                        \A{cc'}{\grg\grg'}(iu) du.
   \label{eq:Edisp33MP}
\end{align}
This is the generalized (distributed) multipole expansion for the three-body non-additive
dispersion energy. 

For systems with large HOMO--LUMO gaps (band gaps in infinite systems) Misquitta \etal 
\cite{MisquittaSSA10} have shown that the non-local polarizabilities decay rapidly
with inter-site separation. The characteristic decay length becomes smaller as the
gap increases. In this case, the non-local polarizabilities can be {\em localized} using 
a multipole expansion \cite{LeSueurS94,LillestolenW07} and we can replace 
$\A{aa'}{\gra\gra'}$ by a local equivalent $\A{a}{\gra\gra'} \delta_{aa'}$ in 
Eq.~\eqref{eq:Edisp33MP} to give:
\begin{align}
  \EdispMP{3}[3](\mbox{loc}) & = 
       + \frac{\hbar}{\pi(\fPIe)^3} \T{ab}{\gra'\grb} \T{bc}{\grb'\grg} \T{ca}{\grg'\gra}
         \nonumber \\
    & ~~~~~ \times ~ \int_0^{\infty} \A{a}{\gra\gra'}(iu) \A{b}{\grb\grb'}(iu) \A{c}{\grg\grg'}(iu) du.
    \label{eq:Edisp33MPloc}
\end{align}
This is the form of the three-body non-additive dispersion energy derived by 
Stogryn \cite{Stogryn71a}, which is valid for large-gap systems only. 
If we retain only the dipole-dipole terms in the Stogryn expression and make the
further assumption that we are dealing with systems of isotropic sites of
(average) polarizability $\bar{\alpha}^{a}$, we can use
$\A{a}{\gra\gra'} \delta_{aa'} = \bar{\alpha}^{a}\delta_{\gra\gra'}$,
and we obtain the Axilrod--Teller--Muto \cite{AxilrodT43,Muto43} triple-dipole term
\cite{AxilrodT43,Muto43}:
\begin{align}
  \EdispMP{3}[3,\mbox{ATM}] & = \sum_{abc} C^{abc}_9 
                                 \frac{1 + 3 \cos \hat{a} \cos \hat{b} \cos \hat{c}}
                                      {(\fPIe)^3 R_{ab}^3 R_{ac}^3 R_{bc}^3}
  \label{eq:ATM}
\end{align}
where the $C^{abc}_9$ dispersion coefficient is defined by
\begin{align}
  C^{abc}_9 = \frac{3\hbar}{\pi} \int_0^{\infty} \bar{\alpha}^{a}(iu) \bar{\alpha}^{b}(iu) 
                                           \bar{\alpha}^{c}(iu) du
\end{align}
and $\hat{a}$ is the angle subtended at site $a$ by unit vectors $\hat{\rr}^{ab}$ and
$\hat{\rr}^{ac}$, with similar definitions for the angles $\hat{b}$ and $\hat{c}$.
This is the more commonly used form of the non-additive dispersion energy, though,
as we see from this derivation, like the Stogryn expression, Eq.~\eqref{eq:ATM}
is valid only for large-gap systems (insulators).

\section{Computational details and results}
\label{sec:results}

\subsection{$\Edisp{3}[3]$ from non-local polarizabilities}
\label{ssec:results-pol}
The na\"{i}ve evaluation of Eq.~\eqref{eq:Edisp33MP} incurs a computational cost that
scales as $\mathcal{O}(n^6 (l+1)^{12} K)$, where $n$ is the number of sites,
$l$ is maximum rank of the polarizability matrix, and $K$ is the number of
quadrature points, typically 10. The
scaling may be improved by calculating and storing the following 
intermediates:
\begin{align}
  \IIu{ab}{\gra\grg} & = \sum_{a',\grb} \Au{aa'}{\gra\grb} \T{ba'}{\grg\grb} 
     \nonumber \\
  \IIu{bc}{\grg\gre} & = \sum_{b',\grd} \Au{bb'}{\grg\grd} \T{cb'}{\gre\grd}
     \nonumber \\
  \IIu{ca}{\gre\gra} & = \sum_{c',\grp} \Au{cc'}{\gre\grp} \T{ac'}{\gra\grp}
\end{align}
The total computational cost of calculating these intermediates is 
$\mathcal{O}(n^3(l+1)^6 K)$. Equation \eqref{eq:Edisp33MP} now takes
the form
\begin{align}
 \EdispMP{3}[3] & = \frac{\hbar}{\pi (\fPIe)^3}
  \int_{0}^{\infty} \IIu{ab}{\gra\grg} \IIu{bc}{\grg\gre} \IIu{ca}{\gre\gra} 
       du \nonumber \\
       & = \frac{\hbar}{\pi (\fPIe)^3} \int_{0}^{\infty} 
          \JJu{ac}{\gra\gre} \IIu{ca}{\gre\gra} du
      \label{eq:3bodydisp-efficient}
\end{align}
where we have defined yet another intermediate
\begin{align}
  \JJu{ac}{\gra\gre} = \sum_{b,\grg} \IIu{ab}{\gra\grg} \IIu{bc}{\grg\gre}
\end{align}
which incurs a computational cost of $\mathcal{O}(n^3 (l+1)^6 K)$.
Equation \eqref{eq:3bodydisp-efficient} is evaluated with a computational
cost of $\mathcal{O}(n^2 (l+1)^4 K)$, so the overall cost of the
calculation is only $\mathcal{O}(4 n^3 (l+1)^6 K)$; a significant improvement
from the na\"{i}ve cost reported above.

We have studied the interactions between two parallel {\it finite} 
\HH{64} chains with bond-alternation parameters $\distortion = 2.0, 1.5$ and $1.0$, 
where \distortion is the ratio of the alternate bond lengths. 
Frequency-dependent polarizability calculations were performed with
coupled Kohn--Sham perturbation theory using the PBE functional and the adiabatic
LDA linear-response kernel with the Sadlej-pVTZ basis set \cite{Sadlej88}.
Calculations on shorter chains indicated that the PBE results were qualitatively the
same as those from the more computationally demanding PBE0 functional.
The Kohn--Sham DFT calculations were performed using the {\sc NWChem} program
\cite{NWChem} and the coupled Kohn--Sham perturbation theory and polarizability
calculations were performed with the \CamCASP program \cite{CamCASP}. 
Dispersion energies were calculated with the {\sc Dispersion} program that 
is available upon request.

The finite hydrogen chains with bond-length alternation is a convenient model
for 1D wires as we can control the metallicity of the system 
using the alternation parameter $\distortion$: with $\distortion=2.0, 1.5$ and
$1.0$, the Kohn--Sham HOMO--LUMO gap of the chain is $7.5, 3.1$ and $1.6$ eV,
respectively, the undistorted chain being the most metallic. 

We have calculated distributed non-local polarizabilities with terms from 
rank $0$ (charge) to $4$ (hexadecapole) using a constrained density-fitting algorithm \cite{MisquittaS06}.
This technique has been demonstrated to result in a compact and accurate description
of the frequency-dependent
polarizabilities, with relatively small charge-flow terms. Furthermore, 
Misquitta \etal \cite{MisquittaSSA10} have demonstrated that these polarizabilities
can accurately model the two-body dispersion energies between hydrogen chains 
for which terms of rank $0$ are sufficient; 
the agreement with non-expanded SAPT(DFT) \Edisp{2} energies being excellent 
even for chain separations as small as $6$ a.u. We expect a similar accuracy for the
three-body non-additive dispersion energy investigated in this paper.

\begin{figure}
  \includegraphics[width=0.5\textwidth,clip]{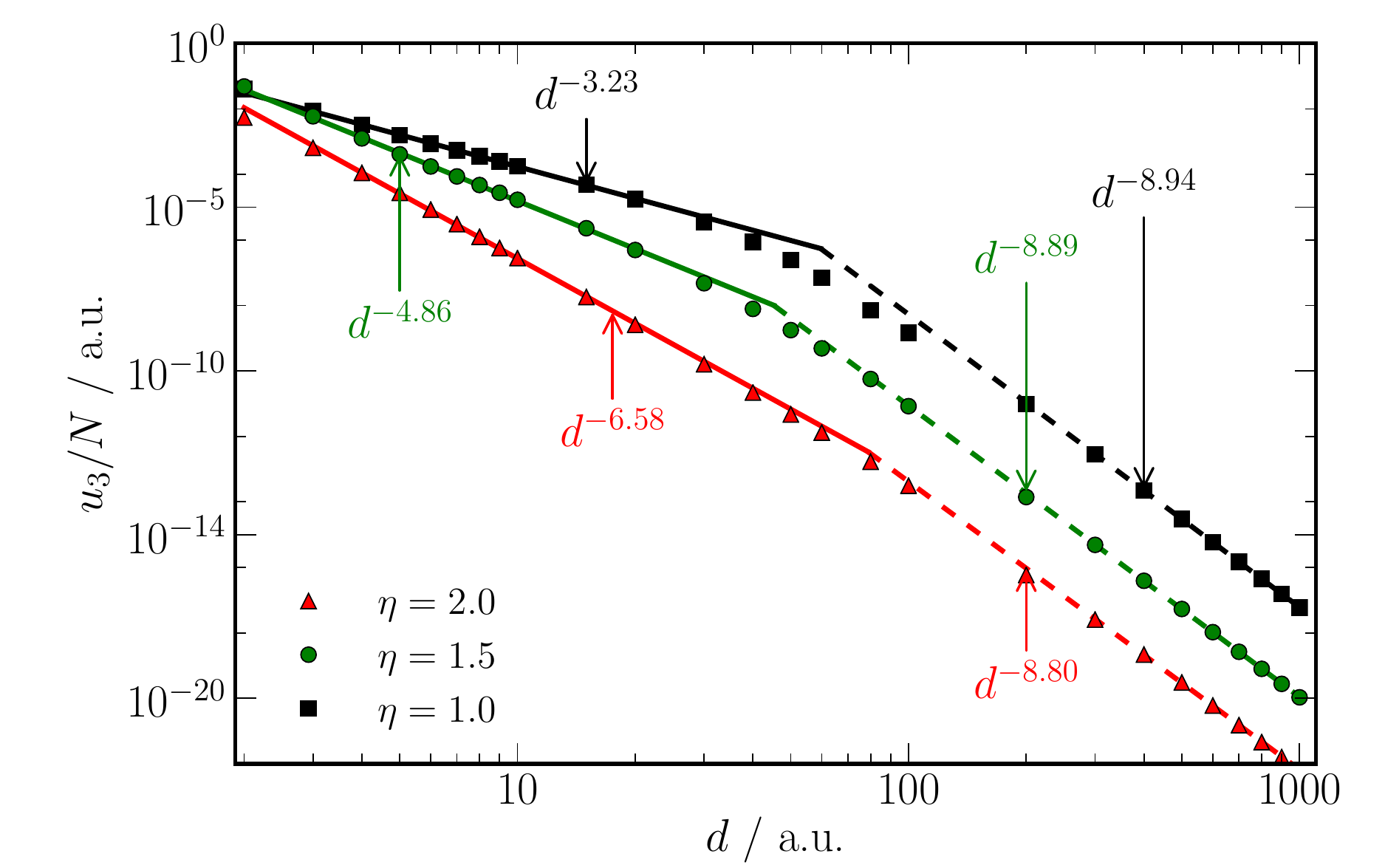}
  \caption[Three-body non-additive dispersion from polarizabilities of \HH{64}-chains]{
  The third-order non-additive dispersion energy calculated using the 
  non-local charge-flow (rank $0$) polarizabilities of \HH{64} chains with bond alternation parameters 
  $\distortion=1$, $1.5$ and $2$. 
  The wires are parallel and arranged in an equilateral triangular configuration
  with side $d$.
  Each set of data is associated with two straight-line fits of the form $\sim d^{-x}$
  to the data in the near (solid lines) and far (dashed lines) regions.
  Broadly, the transition from the short- to long-range behaviour is
  in the region of the intersection of these lines.
  \label{fig:triwire-tri-n64}
  }
\end{figure}
\begin{figure}
  \includegraphics[width=0.5\textwidth,clip]{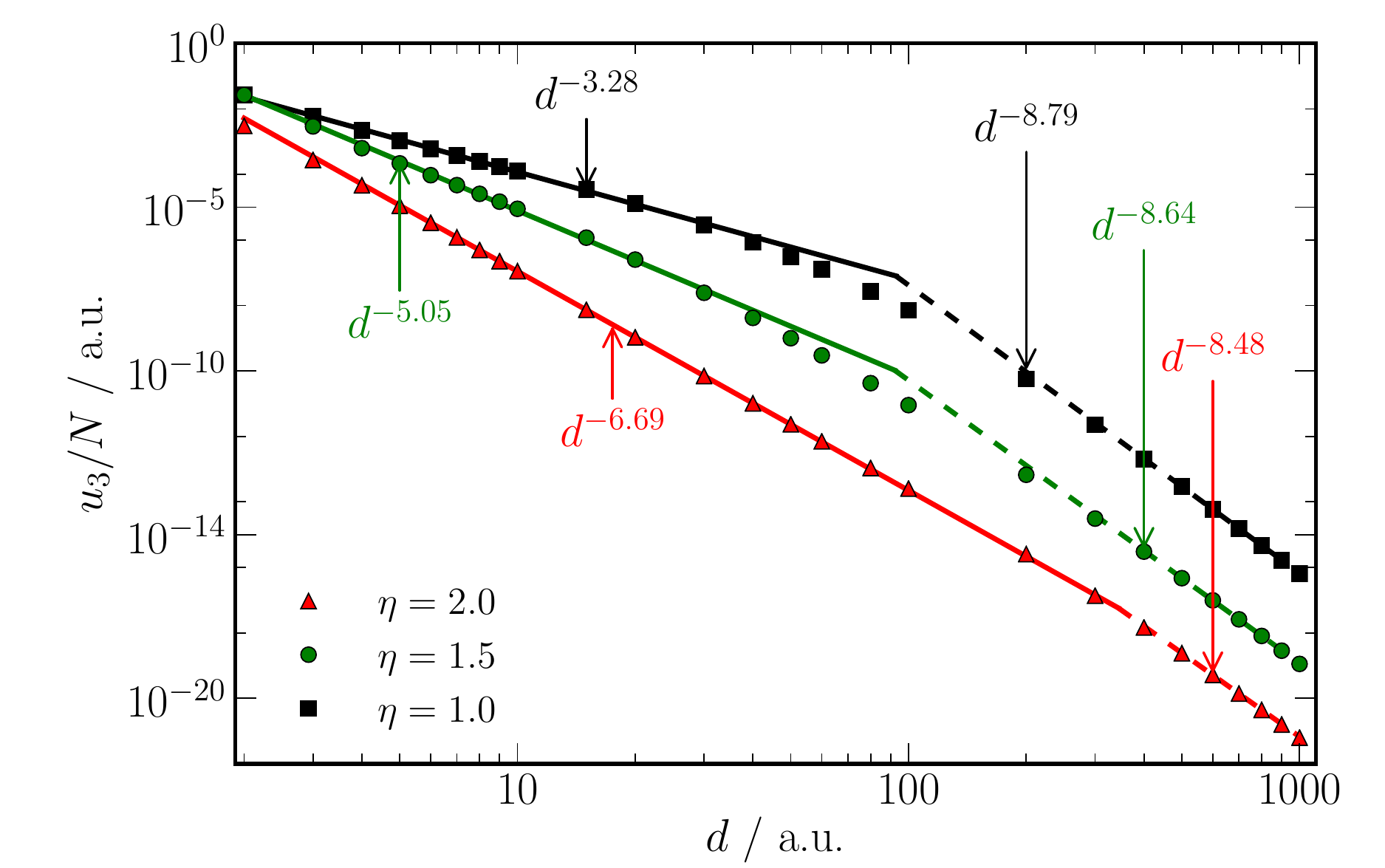}
  \caption[Three-body non-additive dispersion from polarizabilities of \HH{64}-chains]{
  The third-order non-additive dispersion energy calculated using the 
  non-local charge-flow (rank $0$) polarizabilities of \HH{64} chains with bond alternation parameters 
  $\distortion=1$, $1.5$ and $2$. 
  The wires are parallel, coplanar and equally spaced.
  \label{fig:triwire-par-n64}
  }
\end{figure}

In Figs.~\ref{fig:triwire-tri-n64} and \ref{fig:triwire-par-n64} we report 
$\EdispMP{3}[3]$ energies per H$_2$ unit for the equilateral triangular
and coplanar configurations of the \HH{64} trimer. 
The broad features of these figures are:
\begin{itemize}
  \item 
  There is no single power law that fits the data. Instead we have two distinct 
  regions: for separations much larger than the chain length (much greater than 
  70--100 a.u.)
  the non-additive dispersion energy decays as $\sim d^{-9}$, consistent with the
  Axilrod--Teller--Muto expression (Eq.~\eqref{eq:ATM}). This is because
  at such large separations the chains appear to each other as point particles.
  \item
  At sufficiently short separations we see another power-law decay, but with an exponent 
  that varies with the bond alternation, \distortion, of the wire. For the most
  insulating wire with $\distortion=2.0$ the short-separation exponent is relatively
  close to $7$, the value expected from the summation of trimers of atoms,
  while for the most metallic wire with $\distortion=1.0$ the exponent is close
  to $3$.
  \item 
  The non-additive dispersion energy is {\em enhanced} as the degree of metallicity
  increases, and for the most metallic wires is nearly four orders of magnitude larger
  than that for the most insulating wire. 
  \item 
  The charge-flow polarizabilities are responsible for both the change in 
  power-law exponent at short range and the enhancement at long range. 
  Contributions from non-local dipole fluctuations, that is, terms of rank $1$ 
  (not shown in the figures), are insignificant by comparison.
  This was also the observation of 
  Misquitta \etal \cite{MisquittaSSA10} for the two-body dispersion energy.
  \item 
  The Axilrod--Teller--Muto triple dipole expression leads to a favourable
  three-body non-additive dispersion energy for three atoms in a linear 
  configuration. However, for three wires in such a configuration 
  (Fig.~\ref{fig:triwire-par-n64}) the non-additivity is positive, i.e., unfavourable.
\end{itemize}

These observations should perhaps not come as a surprise as they are analogous to
those obtained by Misquitta \etal \cite{MisquittaSSA10} for the two-body dispersion energy 
between 1D wires. However the deviations from the standard picture are much
more dramatic here. In going from the insulating, $\distortion=2.0$, to near-metallic
wire the two-body dispersion exhibits a large-separation enhancement of two orders 
of magnitude compared with four orders for the three-body non-additive dispersion, and 
for small wire separations the power-law changes from $d^{-5}$ to $d^{-2}$ for the
two-body energy while it changes from $d^{-7}$ to $d^{-3}$ for the three-body non-additivity.

\begin{figure}
  \includegraphics[width=0.17\textwidth,clip]{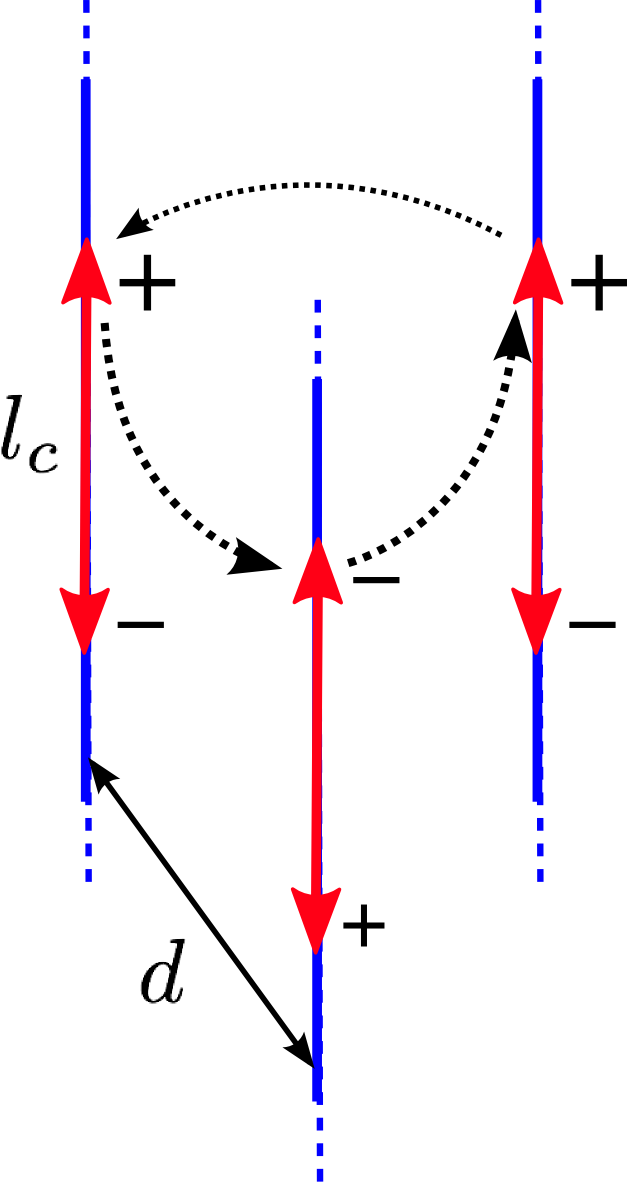}
  \caption[Physical origin of anomalous non-additive three-body dispersion energy
  between three parallel 1-D wires]{
  The anomalous three-body non-additive dispersion interaction between three
  parallel 1D wires (in blue) in an equilateral arrangement can be
  rationalised on the basis of correlations in long-range fluctuations
  (red arrows). Here $d$ is the side of the triangle and $l_c$ is the typical
  correlation length for electronic fluctuations. 
  The spontaneous and induced extended fluctuations are indicated by the
  double-headed arrows, and their signs by the $+\cdots-$ labels.
  \label{fig:edisp33-physics}
  }
\end{figure}

In an analogous manner to the second-order dispersion energy \Edisp{2}, the
anomalous nature of $\Edisp{3}[3]$ can be explained using a simple charge-fluctuation
picture. In Fig.~\ref{fig:edisp33-physics} we depict the
plasmon-like long-range electronic fluctuations in the wires arranged in the
equilateral triangular geometry. The dispersion interaction will be associated
with both local and extended fluctuations. The local fluctuations give rise
to the standard model for $\Edisp{3}[3]$. Here we are concerned with the 
extended, plasmon-like fluctuations of typical length-scale $l_c$,
as depicted in the figure.
An extended $+\cdots-$ spontaneous fluctuation in one wire induces a 
$-\cdots+$ fluctuation in the second, which in turn, induces a $+\cdots-$
fluctuation in the third. The interaction between the first and third will
always be repulsive leading to a positive $\Edisp{3}[3]$ energy.
If the wire separations satisfies $d \lt l_c$, the extended fluctuations
cannot be regarded as dipoles, instead, as shown in 
Fig.~\ref{fig:edisp33-physics}, their interactions are modelled as 
between two trimers of charges resulting from extended charge fluctuations.
Each pair of charges in a trimer interacts as $d^{-1}$, leading to an
effective three-body non-additive dispersion of $\U{3} \sim +d^{-3}$.
For wire separations much larger than $l_c$, the extended 
fluctuations can be modelled as dipoles. Each pair of these dipoles interacts 
as $\pm d^{-3}$, giving rise to a $+d^{-9}$ contribution to the non-additive
dispersion energy.
But all such interactions must be summed over, leading to the effective 
$\U{3} \sim +d^{-7}$ behaviour.
If the wires are finite in extent, we recover the  $\U{3} \sim +d^{-9}$ power law
for separations much larger than the wire length.

It is now well-known that Kohn--Sham time-dependent linear-response theory is not 
quantitatively accurate for heavily delocalized systems, with polarizabilities typically 
overestimated \cite{ChampagnePvGBSS-GRK98,ChampagneMVA95a,ChampagneMVA95b}, and 
hyperpolarizabilities even more so. One may therefore question the validity of
our calculations. We seek, however, a description of the {\em physical} effect 
and make no claims to being quantitatively accurate. We know from the range of calculations
described in the Introduction that our hydrogen chain models are able to 
describe the physics of the two-body dispersion energy between 1D wires
and we see no reason to doubt their validity for trimers of such wires. 
Nevertheless, to remove any possibility of doubt, we have used
QMC techniques to corroborate the results obtained with these
models. 

\subsection{Diffusion Monte Carlo (DMC) calculations}
\label{ssec:results-DMC}

In our DMC calculations we considered parallel biwires and parallel triwires
in an equilateral-triangle configuration with interwire spacing $d$.
Each wire was modelled by a
single-component 1D HEG of density parameter $r_s$ in a cell of length
$L(r_s,N) = 2Nr_s$ subject to periodic boundary conditions, where $N$ is the
number of electrons per wire in the cell.
The electron-electron interaction was modelled
by a 1D Coulomb potential \cite{Saunders94}.  The charge neutrality of each
wire was maintained by 
introducing a uniform line of positive background charge.  To estimate the
asymptotic binding behaviour between long, metallic wires we must have
\begin{equation}
L\left(r_s,N \right) \gg d \gg r_s .
\end{equation}
We chose to work with real wave functions at the $\Gamma$ point of the
simulation-cell Brillouin zone, and the largest systems we considered had $N =
111$ electrons per wire (333 electrons in total for the triwire).  To
investigate finite-size errors we also performed calculations with $N = 5$,
11, 21, and 55 electrons per wire.

We used many-body trial wave functions of Slater-Jastrow-backflow
type.  Each Slater determinant contained plane-wave orbitals of the
form $\exp(ikx)$.  The use of single-component (i.e., fully
spin-polarised) HEGs is justified in Ref.\ \onlinecite{DrummondN07}.  DMC
calculations for strictly 1D systems do not suffer from a fermion sign problem
because the
nodal surface is completely defined by electron coalescence points,
where the trial wave function goes to zero. Our DMC calculations are
therefore essentially exact for the systems studied, although these
systems are finite wires subject to periodic boundary conditions rather
than infinite wires. Electrons in different wires were treated as
distinguishable, so the triwire (biwire) wave function involves the
product of three (two) Slater determinants.  Our Jastrow exponent
\cite{DrummondTN04} was the sum of a two-body function consisting
of an expansion in powers of inter-electron in-wire separation up to
10th order, and a two-body function consisting of a Fourier expansion
with 14 independent reciprocal-lattice points.  These functions
contained optimisable parameters whose values were allowed to differ
for intrawire and interwire electron pairs.

We employed a backflow transformation in which the electron coordinates
in the Slater determinants were replaced by ``quasiparticle coordinates'' that
depend on the positions of all the electrons. We used the two-body backflow
function of
Ref.\ \onlinecite{Lopez-RiosMDTN06}, which consists of an expansion in powers
of inter-electron in-wire separation up to 10th order, again with separate
terms for intrawire and interwire electron pairs.  Backflow functions are
normally used to improve the nodal surfaces of Slater determinants in QMC
trial wave functions \cite{Lopez-RiosMDTN06}.  In the strictly 1D case the
backflow transformation leaves the (already exact) nodal surface unchanged,
but it provides a
compact parameterisation of three-body correlations \cite{LeeD11}.

The values of the optimisable parameters in the Jastrow factor and backflow
function were determined within VMC by minimising the mean absolute deviation
of the local energy from the median local energy \cite{NeedsTDL-R10}. The
optimisations were performed using 32,000 statistically independent electron
configurations to obtain statistical estimators, while 3,200 configurations
were used to determine updates to the parameters
\cite{Trail08a,TrailM10}.

Our DMC calculations were performed with a target population of 1,280
configurations.  The first 500 steps were discarded as equilibration.
To aid comparison of the present results with a previous study \cite{DrummondN07}, we
used the same time steps: 0.04, 0.2, and 2.5 a.u.\ at $r_s = 1$, 3, and 10,
respectively.  These are sufficiently small that the time-step bias in our
results is negligible.
Our QMC calculations were performed using the \textsc{casino} code
\cite{NeedsTDL-R10}.

\subsection{DMC results}

We denote the total energy of the $N$-electron $M$-wire system 
as $E_M$, and the total energy per electron as $e_M$, so $e_1 = E_1/N$.
The parallel 2-wire system has an additional interaction energy $\Delta E_2(d)$,
so the energy per electron is
\begin{equation}
  e_2(d) = (2E_1 + \Delta E_2(d))/2N \equiv e_1 + \U{2}(d),
\end{equation}
consequently the biwire interaction energy per electron $\U{2}(d)$ is
\begin{equation}
  \U{2}(d) = e_2(d) - e_1 .
\end{equation}
Similarly the equilateral-triangle configuration, parallel 3-wire system
has an energy per electron of
\begin{equation}
  e_3(d) = (3E_1 + 3\Delta E_2(d) + \Delta E_3(d)) / 3N \equiv e_1 + 2u_2(d) + u_3(d),
\end{equation}
from which we get the nonadditive contribution to the energy of the triwire system per
electron to be
\begin{equation}
\U{3}(d) = e_3(d) - e_1 - 2 \U{2}(d) = e_3(d) - 2e_2(d) + e_1.
\end{equation}

\begin{figure}[ht]
  \begin{center}
    \includegraphics[width=80mm,angle=0]{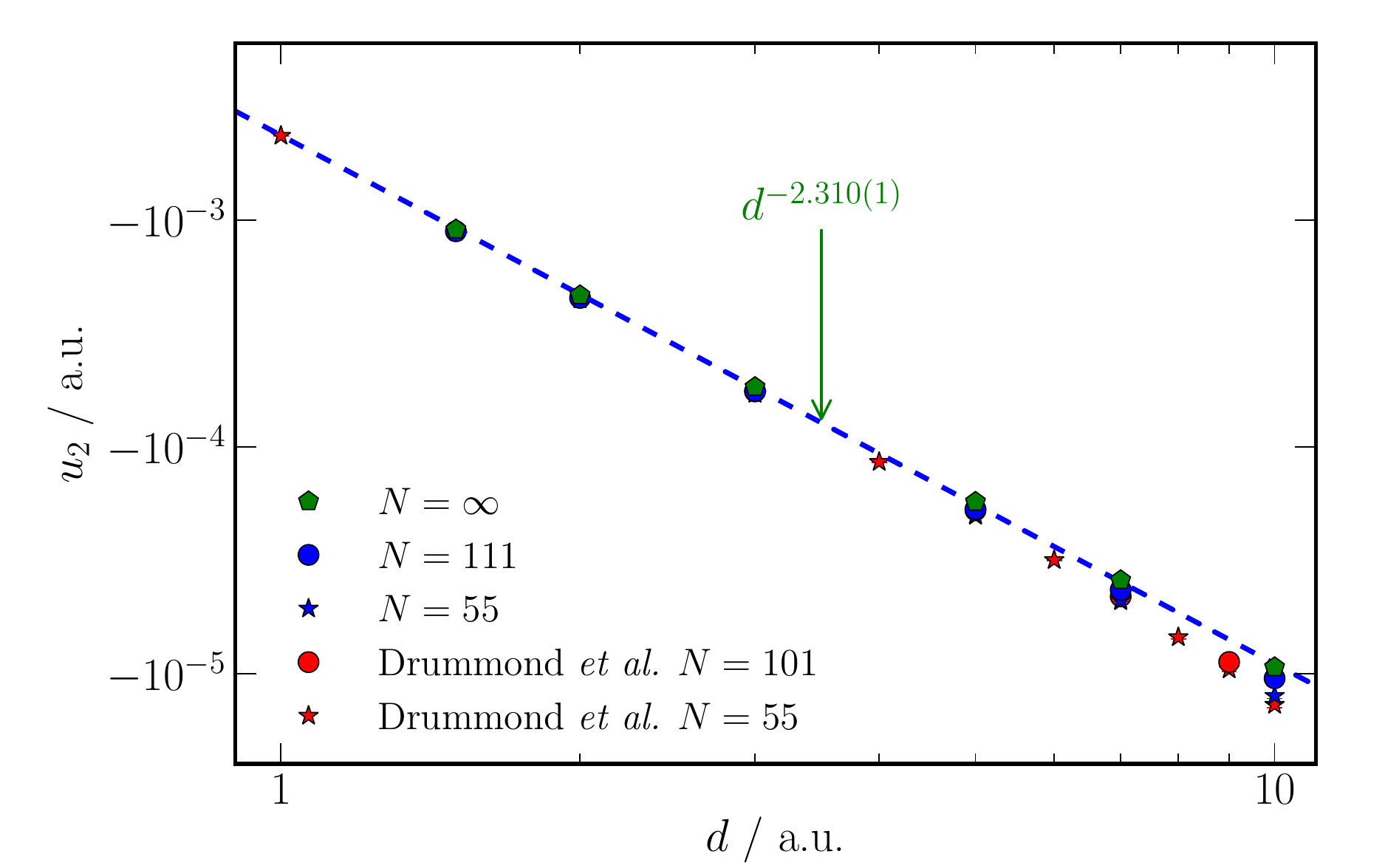}
    \includegraphics[width=80mm,angle=0]{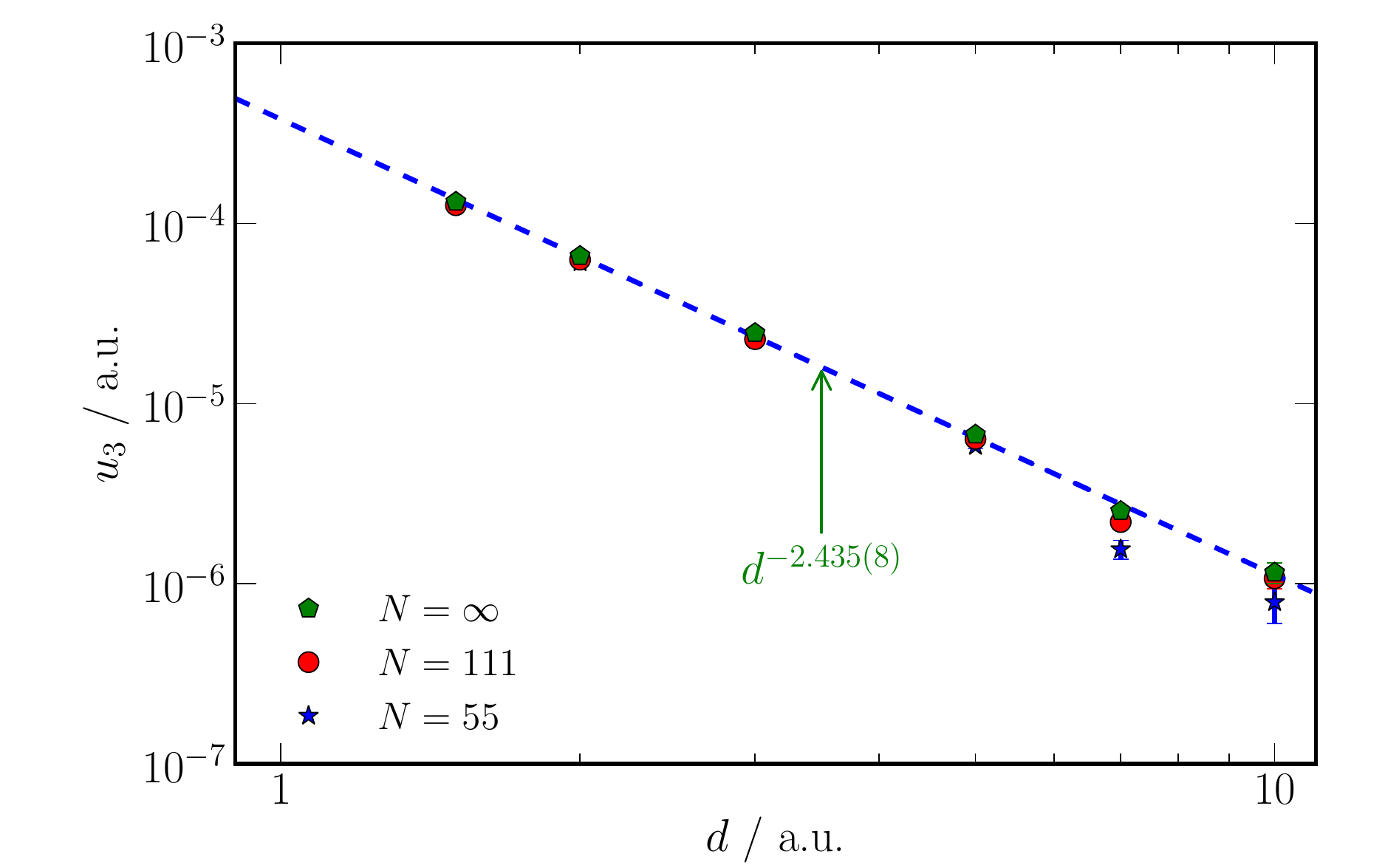}
  \end{center}
\caption{DMC results for the asymptotic behaviour of the biwire interaction \U{2} (left panel)
  and the nonadditive triwire contribution \U{3} (right panel) at $r_s=1$.
}
\label{Fig04}
\end{figure}
\begin{figure}[ht]
  \begin{center}
    \includegraphics[width=80mm,angle=0]{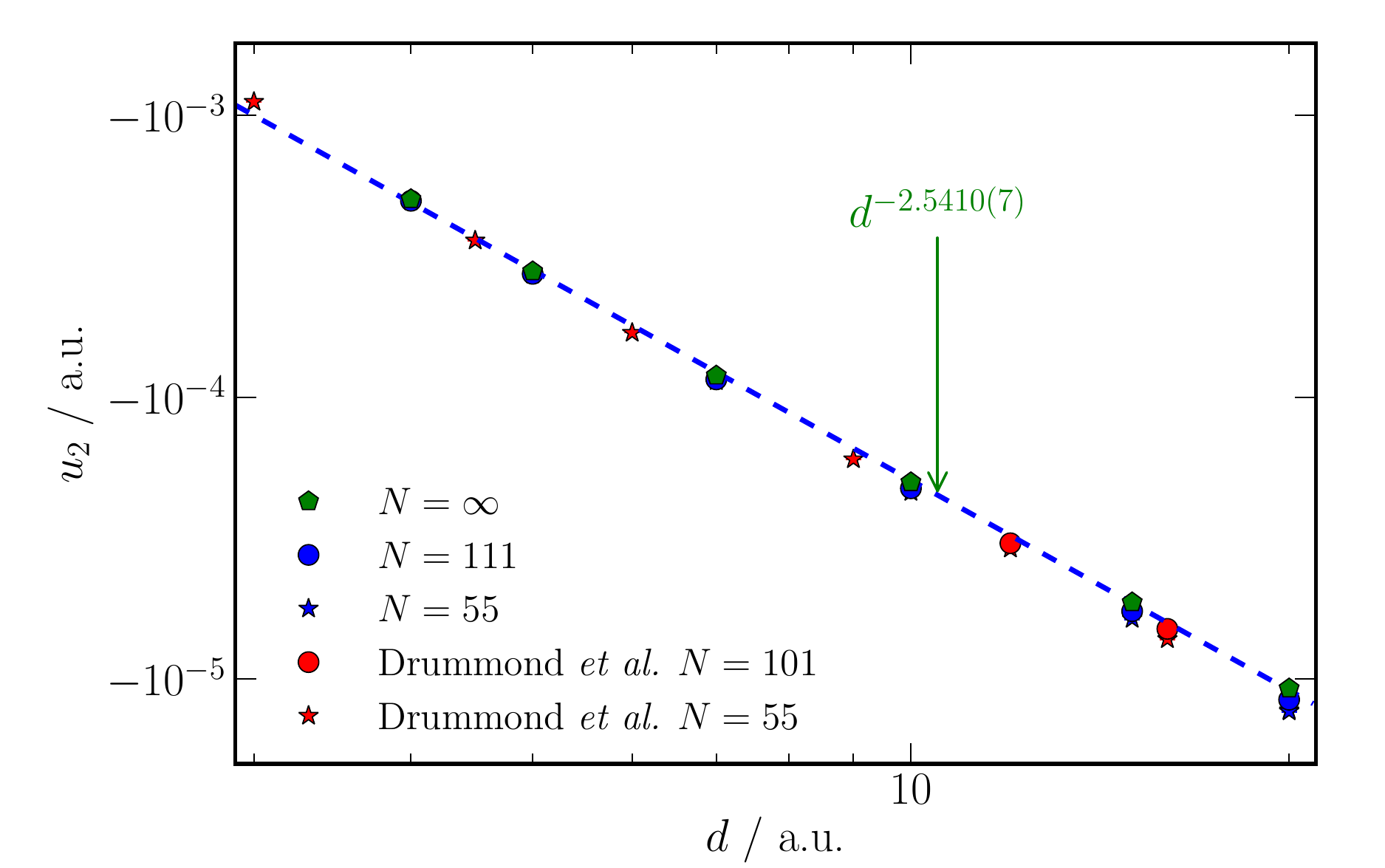}
    \includegraphics[width=80mm,angle=0]{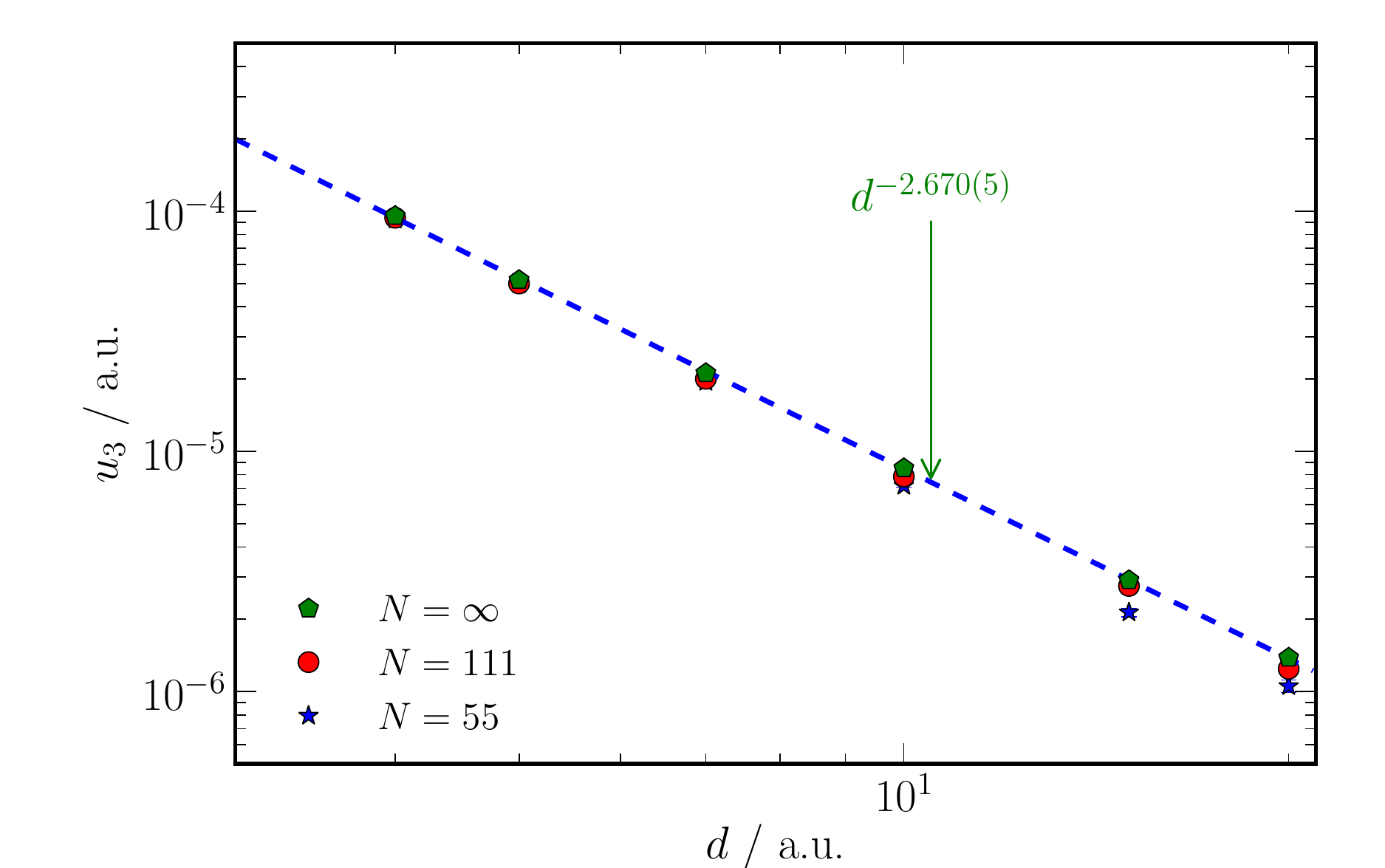}
  \end{center}
\caption{DMC results for the asymptotic behaviour of the biwire interaction \U{2} (left panel) 
and the nonadditive contribution \U{3} (right panel) at $r_s=3$.
}
\label{Fig05}
\end{figure}
\begin{figure}[ht]
  \begin{center}
    \includegraphics[width=80mm,angle=0]{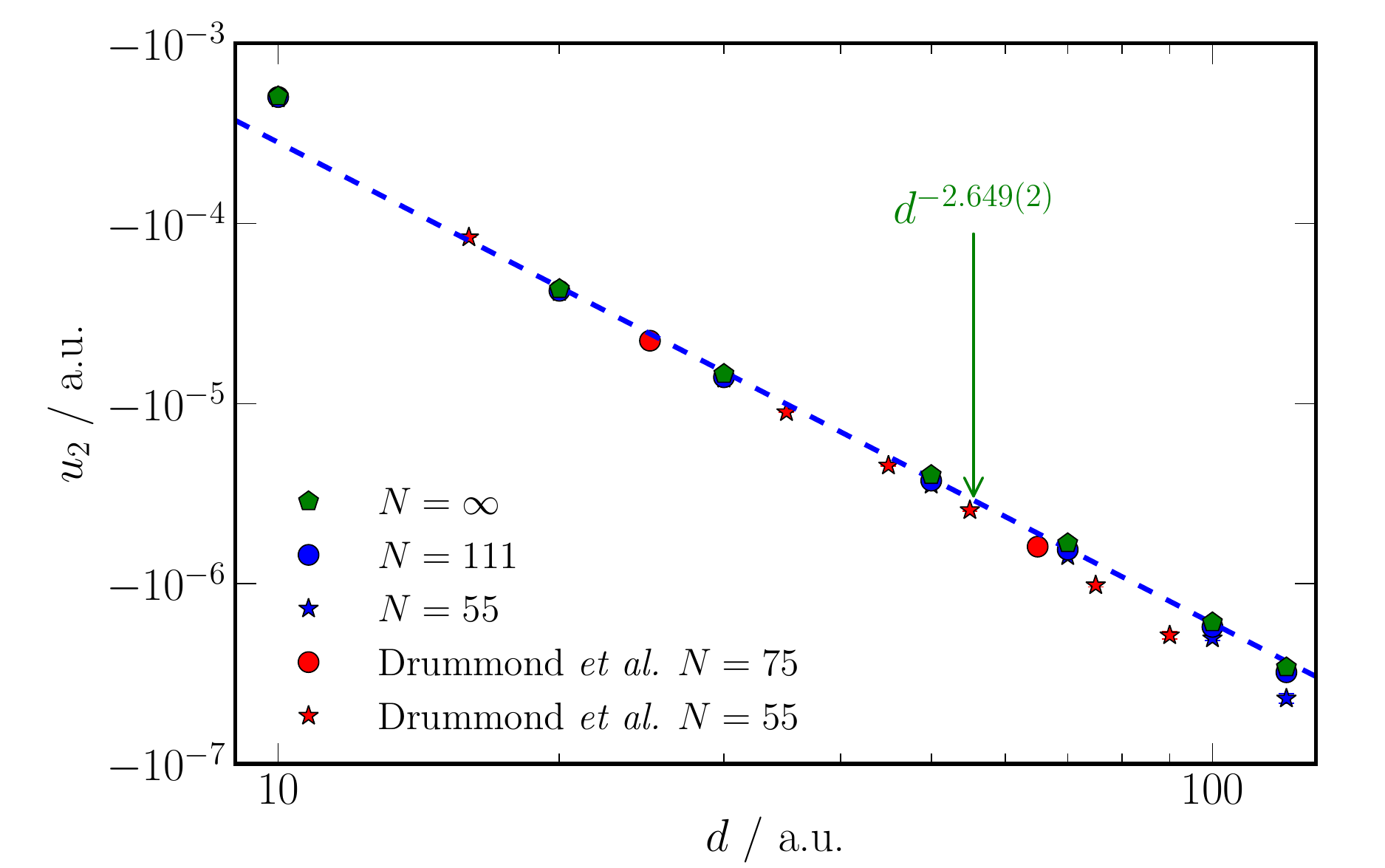}
    \includegraphics[width=80mm,angle=0]{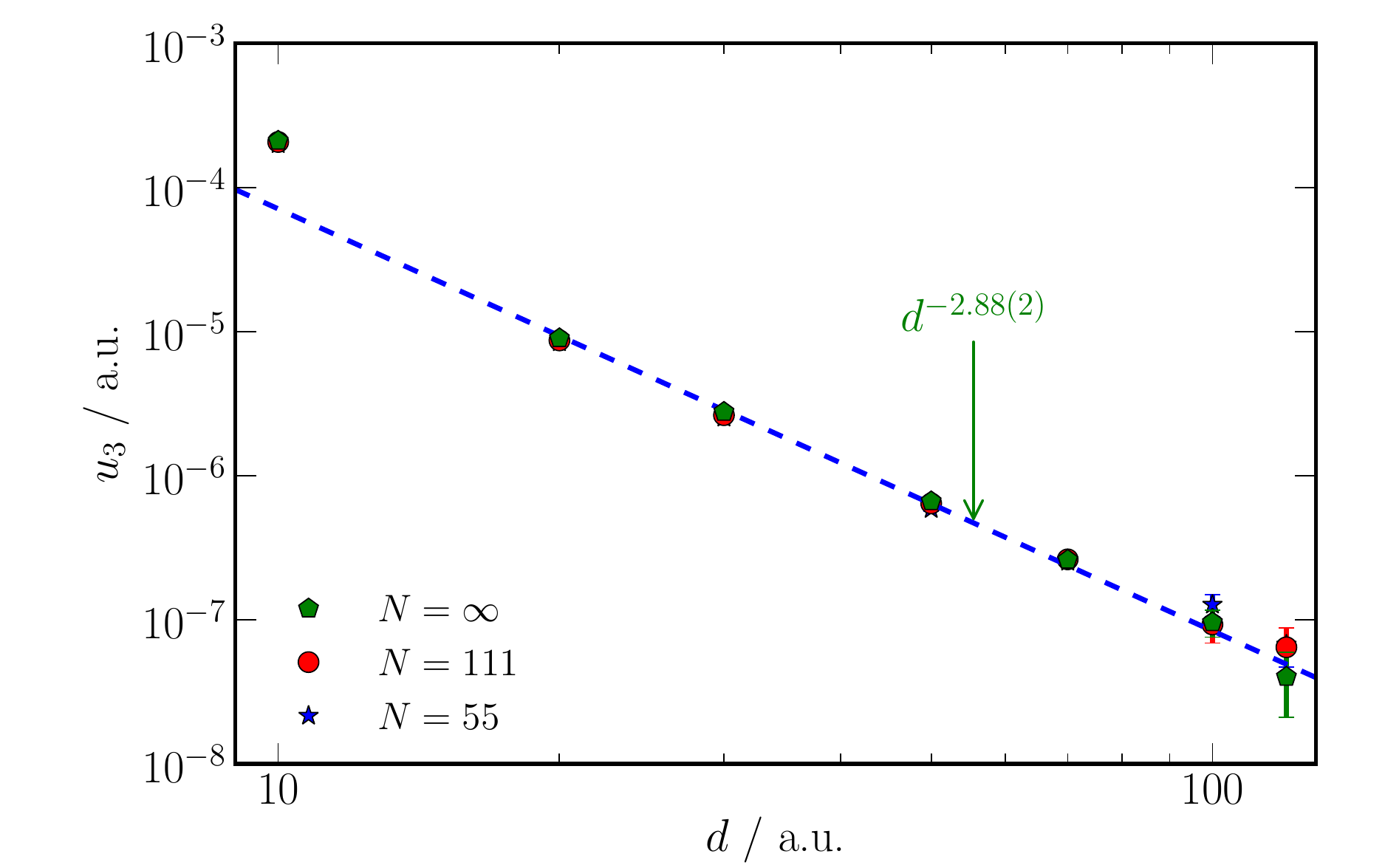}
  \end{center}
\caption{DMC results for the asymptotic behaviour of the biwire interaction \U{2} (left panel) 
and the non-additive contribution \U{3} (right panel) at $r_s=10$.
}
\label{Fig06}
\end{figure}

We fitted
\begin{equation}
\label{eq:power_law}
u(d) = \frac{\exp(C)}{d^\alpha},
\end{equation}
where $C$ and $\alpha$ are fitting parameters, to our DMC results for
$|\U{2}(d)|$ and $|\U{3}(d)|$ (extrapolated to the thermodynamic limit), for
$d$ in the asymptotic regime.
As shown in
Figs.\ \ref{Fig04}--\ref{Fig06}, the asymptotic binding energies $\U{2}(d)$
and $\U{3}(d)$ show power-law behaviour as a function of $d$ at all
densities.





\begin{figure}[ht]
\begin{center}
\includegraphics[width=80mm,angle=0]{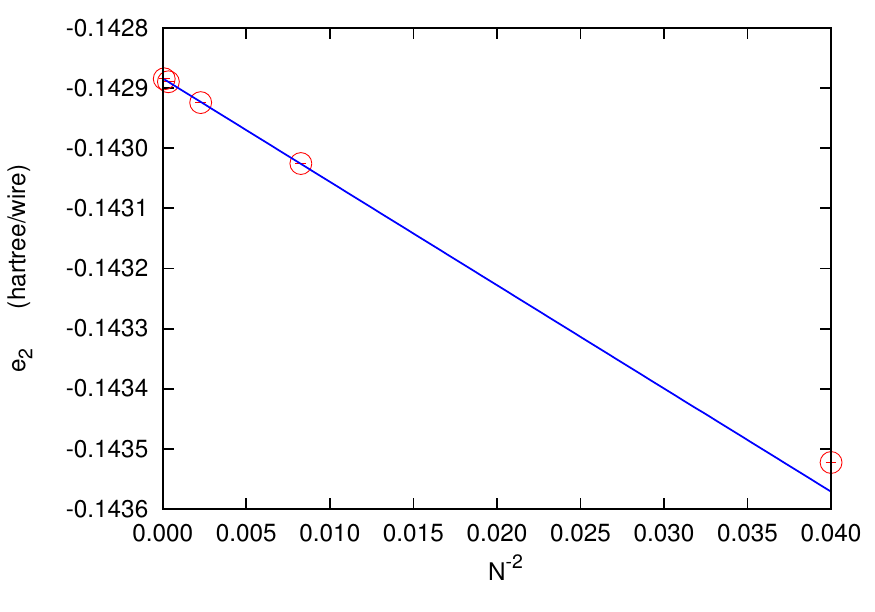}
\includegraphics[width=80mm,angle=0]{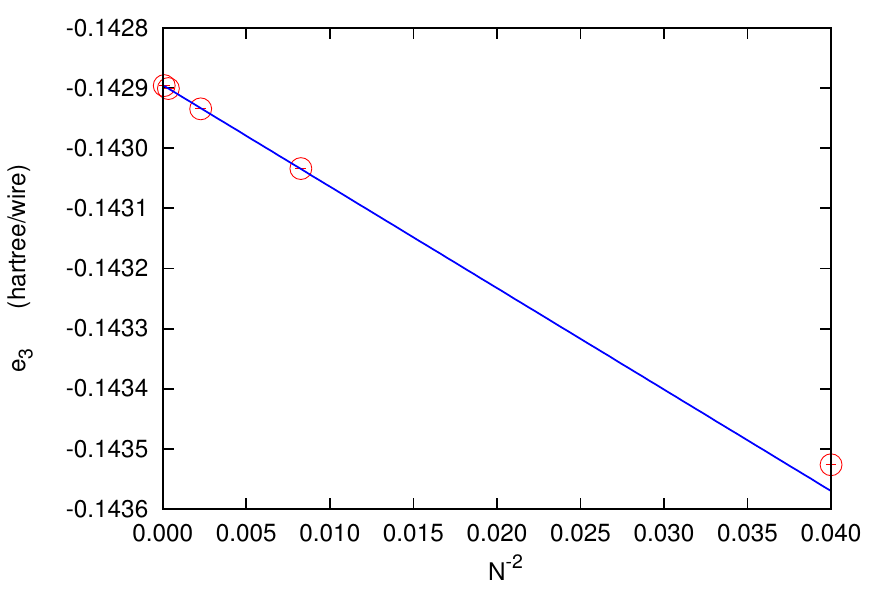}\\
\includegraphics[width=80mm,angle=0]{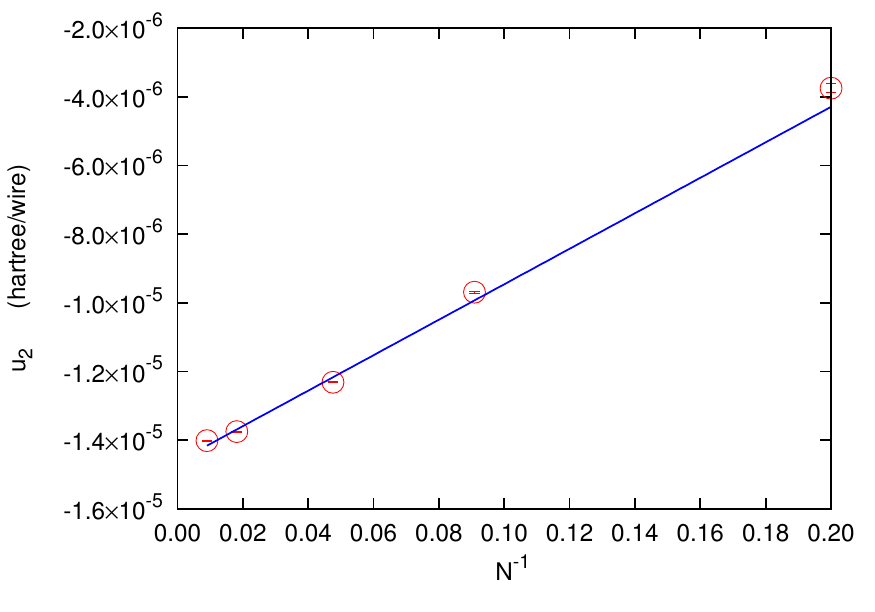}
\includegraphics[width=80mm,angle=0]{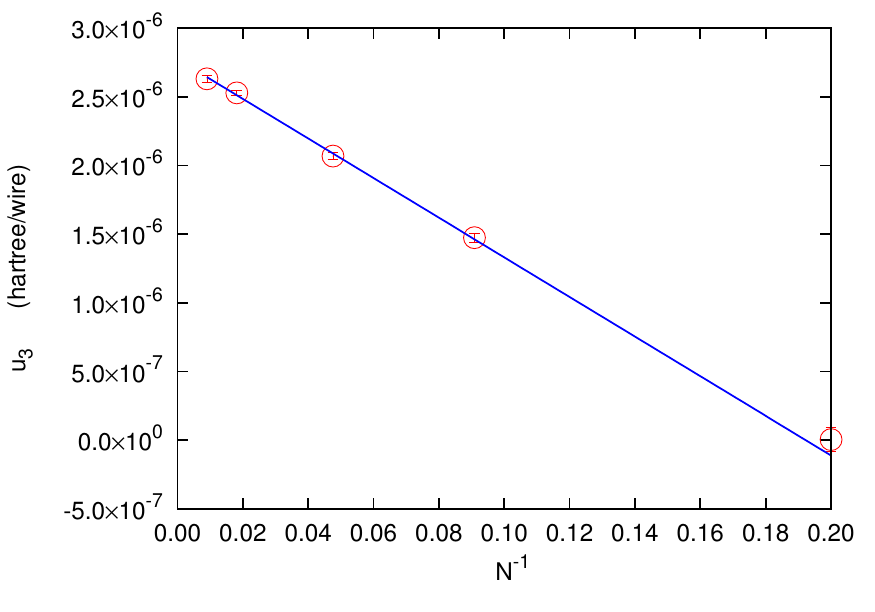}
\end{center}
\caption{DMC results for the $N$-dependence of the total biwire ($e_2$) 
and triwire ($e_3$) energies and interaction energies (\U{2} and \U{3}) 
at $r_s=10$ and at interwire spacing $d=30$ a.u. 
The data at $N=5$ ($1/N=0.2$, $1/N^2 = 0.04$) were excluded from the fits
(solid lines).}
\label{Fig07}
\end{figure}

To estimate the finite-size errors at a given wire separation $d$, we examined
the variation of the energy with the number $N$ of electrons per wire.  It has
recently been reported \cite{LeeD11} that the finite-size error in the total
energy per electron of the 1D HEG scales as
\begin{equation}
e_1(N) = e_1(\infty) + \frac{c}{N^2}, \label{eq:N-2_scaling}
\end{equation}
where $c$ is a constant, over the range of $N$ considered here.  Our results
for $e_2$ and $e_3$, shown in
Fig.\ \ref{Fig07}, are consistent with this dependence.  However, we find that
the interaction energies $u_2$ and $u_3$ at a given $d$ show a more slowly
decaying finite-size error:
\begin{equation}
u_M\left( N \right) = u_M(\infty) + \frac{c^\prime}{N}, \label{eq:fs_extrap}
\end{equation}
where $c^\prime$ is a constant.  Hence Eq.\ (\ref{eq:N-2_scaling}) cannot
give the asymptotic form of the finite-size error in the total energy of a 1D
system in the limit of large $N$.

We have extrapolated the binding-energy data shown in
Figs.\ \ref{Fig04}--\ref{Fig06} to the thermodynamic limit at each $d$ using
Eq.\ (\ref{eq:fs_extrap}).  We have then fitted Eq.\ (\ref{eq:power_law})
to the extrapolated binding-energy data for triwires and
biwires, respectively. The resulting fitting parameters, including the
asymptotic exponents, are given in Table \ref{tab:Fit1}. 


\begin{table}[ht]
  \caption{Values of power-law parameters in Eq.\ (\ref{eq:power_law}) for the two-body
  and three-body energies. }
  \label{tab:Fit1}
  \begin{center}
    \begin{tabular}{l Z{8} Z{8} c Z{8} Z{8} }
      \toprule
                 & \multicolumn{2}{c}{$\U{2} \lt 0$} 
                                           & & \multicolumn{2}{c}{$\U{3} \gt 0$}\\
        \cline{2-3}\cline{5-6}
                 &\multicolumn{1}{c}{$C$}
                              &\multicolumn{1}{c}{$\alpha$}
                                           & & \multicolumn{1}{c}{$C$}
                                                       &\multicolumn{1}{c}{$\alpha$}  \\
      \colrule
        $r_s=1$  & -6.0685(6) & 2.310(1)   & & -7.942(5) & 2.435(8) \\
        $r_s=3$  & -4.084(1)  & 2.5410(7)  & & -5.565(8) & 2.670(5) \\
        $r_s=10$ & -2.114(6)  & 2.649(2)   & & -2.98(5)  & 2.88(2) \\
      \botrule
    \end{tabular}
  \end{center}
\end{table}


\section{Discussion}
\label{sec:discussion}

We have investigated the nature of the non-additive dispersion between three
parallel wires and we have demonstrated that as the HOMO--LUMO gap (band gap
in infinite wires) decreases, the deviations of $\Edisp{3}[3]$ from the
conventional triple-dipole Axilrod--Teller--Muto model increase. These
deviations occur mainly in two ways:
\begin{itemize}
\item 
  For wire separations smaller than the typical electron correlation 
  length, the effective three-body non-additive dispersion behaves
  as $\U{3}(d) \sim d^{-\beta}$, where $\beta \rightarrow 3$ as the 
  HOMO--LUMO gag decreases.
  This power-law arises from the correlations between extended charge
  fluctuations that are associated from the plasmon-like modes in the 
  wires.
  This is a substantially slower decay than the 
  $\U{3}(d) \sim d^{-7}$ behaviour expected from the standard triple-dipole
  summations associated with local dipole fluctuations.
  For finite wires, $\U{3}(d) \sim d^{-9}$ for separations much
  larger than the wire length.
\item
  $\U{3}(d)$ is substantially enhanced as the gap reduces. This is most
  dramatic for large separations, where we observed an enhancement of four
  orders of magnitude for the near-metallic wires compared with the wires with
  the largest HOMO--LUMO gap.
\end{itemize}
These observations are analogous to those obtained by Misquitta 
\etal \cite{MisquittaSSA10} with regard to the second-order dispersion
energy \Edisp{2}, though the effects of metallicity are more dramatic
for the three-body non-additivity.
We have provided a simple physical picture of correlations in extended
charge fluctuations using which both of these observations can be 
understood. 

We have established these results using two techniques: (1) a generalised 
multipole expansion for $\Edisp{3}[3]$ that includes contributions from
charge-flow polarizabilities responsible for the long-wavelength, plasmon-like
fluctuations, and (2) DMC\@. The former has the advantage
that we can directly calculate $\Edisp{3}[3]$, but it is applicable only to 
finite systems with non-zero HOMO--LUMO gaps.
By contrast, DMC 
is applicable to infinite systems (modelled in cells subject to periodic
boundary conditions) with zero gaps, and in principle is
able to describe the third-order correlation energy exactly.
However, like any supermolecular technique, that is, techniques that calculate
the interaction energy from total energy differences, DMC is unable to
separate the two-body energy $\Edisp{3}[2]$ from the three-body 
non-additive dispersion $\Edisp{3}[3]$. 
Nevertheless, there is a consistency in the results from these two methods.
At short range (i.e., at separations less than the correlation length)
the multipole expansion used on trimers of finite \HH{64} chains
yields a power-law of $\U{3}(d) \sim d^{-\beta}$ where $\beta \rightarrow 3^{+}$,
that is, $\beta$ approaches $3$ from above, while in the DMC results,
$\beta \rightarrow 3^{-}$ as $r_s$ increases. For small $r_s$ the exponent is
significantly smaller than $3$. This could be because of finite-size effects,
contributions from $\Edisp{3}[2]$, 
or it could be a genuine effect not captured by the multipole expansion.

The increased effect of the plasmon-like, charge-flow fluctuations on
$\Edisp{3}[3]$ compared with $\Edisp{2}$ is related to the long range
of these fields produced by the fluctuations. The dipole fluctuations
in insulators result in electric fields that behave as $r^{-3}$; 
a rapid decay compared with the $r^{-1}$ behaviour of the electric
fields from the plasmon-type fluctuations. Consequently we expect
the many-body expansion to be slowly convergent for conglomerates of
low-dimensional semi-metallic systems. As we have demonstrated, the
three-body non-additivity quenches the already enhanced two-body
dispersion. Likewise, by extending our physical model for these 
anomalous dispersion effects, we expect that the four-body non-additivity
will be attractive and decay as $-d^{-4}$ for 1D metallic systems,
and will consequently quench the three-body non-additivity.

The slow decay and alternating signs of the $N$-body non-additive dispersion
suggests that the many-body expansion may not be a useful way of modelling
the dispersion interaction in, say, a bundle of 1D semi-metallic wires.
An alternative may be a generalisation of the self-consistent polarization model
proposed by Silberstein \cite{Silberstein17} and Applequist \cite{ApplequistCF72},
and recently significantly developed by Tkatchenko \etal \cite{TkatchenkoDiSCS12}.
However, models such as these would have to be modified to include the
charge-flow polarizabilities to be able to describe the metallic
effects described in this article.

For finite molecular systems, the changes in power-law described here are,
to an extent, of academic interest only. In practice, subtle power-law changes
in the dispersion interaction can be easily masked by the other, often larger,
components of the interaction energy, particularly the first-order 
electrostatic energy. While this may be the case, it is the second
effect---the enhancement of the dispersion energy that arises from the
plasmon-like modes---that may have a perceptible effect. The long-wavelength
fluctuations cause an enhancement of the {\em effective} two- and three-body 
dispersion coefficients.
We believe that this effect, which is captured by techniques such as
the Williams--Stone--Misquitta method\cite{MisquittaS08a,MisquittaSP08},
may prove significant even for relatively small molecular systems.
We are currently working to investigate this phenomenon.

\section{Acknowledgments}

Financial support was provided by the UK Engineering and Physical Sciences
Research Council (EPSRC)\@. 
Part of the computations have been performed using the K computer at 
Advanced Institute for Computational Science, RIKEN. R.M. is grateful for financial support
from KAKENHI grants (23104714, 22104011, and 25600156), and from the Tokuyama Science Foundation. 

\newcommand{\JCP}[0]{J. Chem. Phys.\ }
\newcommand{\JPCA}[0]{J. Phys. Chem. A\ }
\newcommand{\JPCB}[0]{J. Phys. Chem. B\ }
\newcommand{\JPCC}[0]{J. Phys. Chem. C\ }
\newcommand{\JPC}[0]{J. Phys. Chem.\ }
\newcommand{\JCTC}[0]{J. Chem. Theory Comput.\ }
\newcommand{\IJQC}[0]{Int. J. Quantum Chem.\ }
\newcommand{\CPL}[0]{Chem. Phys. Lett.\ }
\newcommand{\TCA}[0]{Theor. Chim. Acta\ }
\newcommand{\PR}[0]{Phys. Rev.\ }
\newcommand{\PRA}[0]{Phys. Rev. A\ }
\newcommand{\PRB}[0]{Phys. Rev. B\ }
\newcommand{\PRE}[0]{Phys. Rev. E\ }
\newcommand{\PRL}[0]{Phys. Rev. Lett.\ }
\newcommand{\CR}[0]{Chem. Rev.\ }
\newcommand{\ChemRev}[0]{Chem. Rev.\ }
\newcommand{\NuP}[0]{Nucl. Phys.\ }
\newcommand{\MolP}[0]{Mol. Phys.\ }
\newcommand{\AdQC}[0]{Adv. Quantum Chem.\ }
\newcommand{\CPC}[0]{Comput. Phys. Commun.\ }
\newcommand{\JMS}[0]{J. Mol. Struct.\ }
\newcommand{\CJC}[0]{Can. J. Chem.\ }
\newcommand{\CP}[0]{Chem. Phys.\ }
\newcommand{\JPB}[0]{J. Phys. B: At. Mol. Opt. Phys.\ }
\newcommand{\PCCP}[0]{Phys. Chem. Chem. Phys.\ }
\newcommand{\JCC}[0]{J. Comp. Chem.\ }
\newcommand{\JACS}[0]{J. Am. Chem. Soc.\ }
\newcommand{\AngChemInt}[0]{Angew. Chem. Int. Ed.\ }
\newcommand{\AngChem}[0]{Angew. Chem.\ }
\newcommand{\ActCrysB}[0]{Acta Cryst. B\ }
\newcommand{\IRPC}[0]{Int. Revs. Phys. Chem.\ }
\newcommand{\PNAS}[0]{Proc. Natl. Acad. Sci.\ }
\newcommand{\PRSLA}[0]{Proc. R. Soc. Lond. A\ }
\newcommand{\CEC}[0]{CrystEngComm.\ }
\newcommand{\CGD}[0]{Cryst. Growth Des.\ }
\newcommand{\NJC}[0]{New J. Chem.\ }
\newcommand{\ChemPhysChem}[0]{ChemPhysChem\ }
\newcommand{\APL}[0]{Appl. Phys. Lett.\ }
\newcommand{\ChemComm}[0]{Chem. Commun.\ }
\newcommand{\RevModPhys}[0]{Rev. Mod. Phys.\ }
\newcommand{\AccChemRes}[0]{Acc. Chem. Res.\ }
\newcommand{\SurfSciLett}[0]{Surf. Sci. Lett.\ }
\newcommand{\JPhysCondMat}[0]{J. Phys.: Condens. Matter\ }
\newcommand{\Nature}[0]{Nature\ }
\newcommand{\NatureMat}[0]{Nature Materials\ }
\newcommand{\NaturePhy}[0]{Nature Physics\ }
\newcommand{\NatureComms}[0]{Nature Communications\ }
\newcommand{\JMathChem}[0]{J. Math. Chem.\ }
\newcommand{\PhysLett}[0]{Phys. Lett.\ }


\begin{thebibliography}{41}
\expandafter\ifx\csname natexlab\endcsname\relax\def\natexlab#1{#1}\fi
\expandafter\ifx\csname bibnamefont\endcsname\relax
  \def\bibnamefont#1{#1}\fi
\expandafter\ifx\csname bibfnamefont\endcsname\relax
  \def\bibfnamefont#1{#1}\fi
\expandafter\ifx\csname citenamefont\endcsname\relax
  \def\citenamefont#1{#1}\fi
\expandafter\ifx\csname url\endcsname\relax
  \def\url#1{\texttt{#1}}\fi
\expandafter\ifx\csname urlprefix\endcsname\relax\def\urlprefix{URL }\fi
\providecommand{\bibinfo}[2]{#2}
\providecommand{\eprint}[2][]{\url{#2}}

\bibitem[{\citenamefont{Spencer}(2009)}]{Spencer09:thesis}
\bibinfo{author}{\bibfnamefont{J.}~\bibnamefont{Spencer}}, Ph.D. thesis,
  \bibinfo{school}{St. Catharine's College, University of Cambridge}
  (\bibinfo{year}{2009}).

\bibitem[{\citenamefont{Misquitta et~al.}(2010)\citenamefont{Misquitta,
  Spencer, Stone, and Alavi}}]{MisquittaSSA10}
\bibinfo{author}{\bibfnamefont{A.~J.} \bibnamefont{Misquitta}},
  \bibinfo{author}{\bibfnamefont{J.}~\bibnamefont{Spencer}},
  \bibinfo{author}{\bibfnamefont{A.~J.} \bibnamefont{Stone}}, \bibnamefont{and}
  \bibinfo{author}{\bibfnamefont{A.}~\bibnamefont{Alavi}},
  \bibinfo{journal}{\PRB} \textbf{\bibinfo{volume}{82}},
  \bibinfo{pages}{075312} (\bibinfo{year}{2010}).

\bibitem[{\citenamefont{Drummond and Needs}(2007)}]{DrummondN07}
\bibinfo{author}{\bibfnamefont{N.~D.} \bibnamefont{Drummond}} \bibnamefont{and}
  \bibinfo{author}{\bibfnamefont{R.~J.} \bibnamefont{Needs}},
  \bibinfo{journal}{\PRL} \textbf{\bibinfo{volume}{99}},
  \bibinfo{pages}{166401(4)} (\bibinfo{year}{2007}).

\bibitem[{\citenamefont{Dobson et~al.}(2006)\citenamefont{Dobson, White, and
  Rubio}}]{DobsonWR06}
\bibinfo{author}{\bibfnamefont{J.~F.} \bibnamefont{Dobson}},
  \bibinfo{author}{\bibfnamefont{A.}~\bibnamefont{White}}, \bibnamefont{and}
  \bibinfo{author}{\bibfnamefont{A.}~\bibnamefont{Rubio}},
  \bibinfo{journal}{\PRL} \textbf{\bibinfo{volume}{96}},
  \bibinfo{pages}{073201(4)} (\bibinfo{year}{2006}).

\bibitem[{\citenamefont{Stone}(2013)}]{Stone:book:13}
\bibinfo{author}{\bibfnamefont{A.~J.} \bibnamefont{Stone}},
  \emph{\bibinfo{title}{The Theory of Intermolecular Forces}}
  (\bibinfo{publisher}{Oxford University Press, Oxford}, \bibinfo{year}{2013}),
  \bibinfo{edition}{2nd} ed.

\bibitem[{\citenamefont{Kaplan}(2005)}]{Kaplan05:book}
\bibinfo{author}{\bibfnamefont{I.~G.} \bibnamefont{Kaplan}},
  \emph{\bibinfo{title}{Intermolecular Interactions}}
  (\bibinfo{publisher}{Wiley}, \bibinfo{year}{2005}), \bibinfo{edition}{2nd}
  ed.

\bibitem[{\citenamefont{Foulkes et~al.}(2001)\citenamefont{Foulkes, Mitas,
  Needs, and Rajagopal}}]{FoulkesMNR01}
\bibinfo{author}{\bibfnamefont{W.~M.~C.} \bibnamefont{Foulkes}},
  \bibinfo{author}{\bibfnamefont{L.}~\bibnamefont{Mitas}},
  \bibinfo{author}{\bibfnamefont{R.~J.} \bibnamefont{Needs}}, \bibnamefont{and}
  \bibinfo{author}{\bibfnamefont{G.}~\bibnamefont{Rajagopal}},
  \bibinfo{journal}{Rev. Mod. Phys.} \textbf{\bibinfo{volume}{73}},
  \bibinfo{pages}{33} (\bibinfo{year}{2001}),
  \urlprefix\url{http://link.aps.org/doi/10.1103/RevModPhys.73.33}.

\bibitem[{\citenamefont{Dobson}(2007)}]{Dobson07a}
\bibinfo{author}{\bibfnamefont{J.~F.} \bibnamefont{Dobson}},
  \bibinfo{journal}{Surface Science} \textbf{\bibinfo{volume}{601}},
  \bibinfo{pages}{5667} (\bibinfo{year}{2007}).

\bibitem[{\citenamefont{Dobson et~al.}(2001)\citenamefont{Dobson, McLennan,
  Rubio, Wang, Gould, Le, and Dinte}}]{DobsonMcLRWGLD01}
\bibinfo{author}{\bibfnamefont{J.~F.} \bibnamefont{Dobson}},
  \bibinfo{author}{\bibfnamefont{K.}~\bibnamefont{McLennan}},
  \bibinfo{author}{\bibfnamefont{A.}~\bibnamefont{Rubio}},
  \bibinfo{author}{\bibfnamefont{J.}~\bibnamefont{Wang}},
  \bibinfo{author}{\bibfnamefont{T.}~\bibnamefont{Gould}},
  \bibinfo{author}{\bibfnamefont{H.~M.} \bibnamefont{Le}}, \bibnamefont{and}
  \bibinfo{author}{\bibfnamefont{B.~P.} \bibnamefont{Dinte}},
  \bibinfo{journal}{Aust. J. Chem.} \textbf{\bibinfo{volume}{54}},
  \bibinfo{pages}{513} (\bibinfo{year}{2001}).

\bibitem[{\citenamefont{Coulson and Davies}(1952)}]{CoulsonD52}
\bibinfo{author}{\bibfnamefont{C.~A.} \bibnamefont{Coulson}} \bibnamefont{and}
  \bibinfo{author}{\bibfnamefont{P.~L.} \bibnamefont{Davies}},
  \bibinfo{journal}{Trans. Faraday Soc.} \textbf{\bibinfo{volume}{48}},
  \bibinfo{pages}{777 } (\bibinfo{year}{1952}).

\bibitem[{\citenamefont{Longuet-Higgins and Salem}(1961)}]{Longuet-HigginsS61}
\bibinfo{author}{\bibfnamefont{H.~C.} \bibnamefont{Longuet-Higgins}}
  \bibnamefont{and} \bibinfo{author}{\bibfnamefont{L.}~\bibnamefont{Salem}},
  \bibinfo{journal}{Proc. R. Soc. A} \textbf{\bibinfo{volume}{259}},
  \bibinfo{pages}{433} (\bibinfo{year}{1961}).

\bibitem[{\citenamefont{Chang et~al.}(1971)\citenamefont{Chang, Cooper,
  Drummond, and Young}}]{ChangCDY71}
\bibinfo{author}{\bibfnamefont{D.~B.} \bibnamefont{Chang}},
  \bibinfo{author}{\bibfnamefont{R.~L.} \bibnamefont{Cooper}},
  \bibinfo{author}{\bibfnamefont{J.~E.} \bibnamefont{Drummond}},
  \bibnamefont{and} \bibinfo{author}{\bibfnamefont{A.~C.} \bibnamefont{Young}},
  \bibinfo{journal}{\PhysLett} \textbf{\bibinfo{volume}{37A}},
  \bibinfo{pages}{311} (\bibinfo{year}{1971}).

\bibitem[{\citenamefont{Gobre and Tkatchenko}(2013)}]{GobreT13}
\bibinfo{author}{\bibfnamefont{V.~V.} \bibnamefont{Gobre}} \bibnamefont{and}
  \bibinfo{author}{\bibfnamefont{A.}~\bibnamefont{Tkatchenko}},
  \bibinfo{journal}{\NatureComms} \textbf{\bibinfo{volume}{4}},
  \bibinfo{pages}{2341} (\bibinfo{year}{2013}).

\bibitem[{\citenamefont{Misquitta and Stone}(2008)}]{MisquittaS08a}
\bibinfo{author}{\bibfnamefont{A.~J.} \bibnamefont{Misquitta}}
  \bibnamefont{and} \bibinfo{author}{\bibfnamefont{A.~J.} \bibnamefont{Stone}},
  \bibinfo{journal}{\JCTC} \textbf{\bibinfo{volume}{4}}, \bibinfo{pages}{7}
  (\bibinfo{year}{2008}).

\bibitem[{\citenamefont{Misquitta et~al.}(2008)\citenamefont{Misquitta, Stone,
  and Price}}]{MisquittaSP08}
\bibinfo{author}{\bibfnamefont{A.~J.} \bibnamefont{Misquitta}},
  \bibinfo{author}{\bibfnamefont{A.~J.} \bibnamefont{Stone}}, \bibnamefont{and}
  \bibinfo{author}{\bibfnamefont{S.~L.} \bibnamefont{Price}},
  \bibinfo{journal}{\JCTC} \textbf{\bibinfo{volume}{4}}, \bibinfo{pages}{19}
  (\bibinfo{year}{2008}).

\bibitem[{\citenamefont{Angyan}(2009)}]{Angyan09a}
\bibinfo{author}{\bibfnamefont{J.~G.} \bibnamefont{Angyan}},
  \bibinfo{journal}{\IJQC} \textbf{\bibinfo{volume}{109}},
  \bibinfo{pages}{2340} (\bibinfo{year}{2009}).

\bibitem[{\citenamefont{Parsegian}(2005)}]{Parsegian05:book}
\bibinfo{author}{\bibfnamefont{V.~A.} \bibnamefont{Parsegian}},
  \emph{\bibinfo{title}{Van der Waals Forces: A Handbook for Biologists,
  Chemists, Engineers, and Physicists}} (\bibinfo{publisher}{Cambridge
  University Press}, \bibinfo{year}{2005}).

\bibitem[{\citenamefont{Stogryn}(1971)}]{Stogryn71a}
\bibinfo{author}{\bibfnamefont{D.~E.} \bibnamefont{Stogryn}},
  \bibinfo{journal}{\MolP} \textbf{\bibinfo{volume}{22}}, \bibinfo{pages}{81}
  (\bibinfo{year}{1971}).

\bibitem[{\citenamefont{Axilrod and Teller}(1943)}]{AxilrodT43}
\bibinfo{author}{\bibfnamefont{P.~M.} \bibnamefont{Axilrod}} \bibnamefont{and}
  \bibinfo{author}{\bibfnamefont{E.}~\bibnamefont{Teller}},
  \bibinfo{journal}{\JCP} \textbf{\bibinfo{volume}{11}}, \bibinfo{pages}{299}
  (\bibinfo{year}{1943}).

\bibitem[{\citenamefont{Muto}(1943)}]{Muto43}
\bibinfo{author}{\bibfnamefont{Y.}~\bibnamefont{Muto}}, \bibinfo{journal}{Proc.
  Phys.-Math. Soc. Japan} \textbf{\bibinfo{volume}{17}}, \bibinfo{pages}{629}
  (\bibinfo{year}{1943}).

\bibitem[{\citenamefont{Longuet-Higgins}(1965)}]{Longuet-Higgins65}
\bibinfo{author}{\bibfnamefont{H.~C.} \bibnamefont{Longuet-Higgins}},
  \bibinfo{journal}{Disc. Faraday Soc.} \textbf{\bibinfo{volume}{40}},
  \bibinfo{pages}{7} (\bibinfo{year}{1965}), \bibinfo{note}{spiers Memorial
  Lecture}.

\bibitem[{\citenamefont{Stone}(1985)}]{Stone85}
\bibinfo{author}{\bibfnamefont{A.~J.} \bibnamefont{Stone}},
  \bibinfo{journal}{\MolP} \textbf{\bibinfo{volume}{56}}, \bibinfo{pages}{1065}
  (\bibinfo{year}{1985}).

\bibitem[{\citenamefont{Misquitta and Stone}(2006)}]{MisquittaS06}
\bibinfo{author}{\bibfnamefont{A.~J.} \bibnamefont{Misquitta}}
  \bibnamefont{and} \bibinfo{author}{\bibfnamefont{A.~J.} \bibnamefont{Stone}},
  \bibinfo{journal}{\JCP} \textbf{\bibinfo{volume}{124}},
  \bibinfo{pages}{024111} (\bibinfo{year}{2006}).

\bibitem[{\citenamefont{Le~Sueur and Stone}(1994)}]{LeSueurS94}
\bibinfo{author}{\bibfnamefont{C.~R.} \bibnamefont{Le~Sueur}} \bibnamefont{and}
  \bibinfo{author}{\bibfnamefont{A.~J.} \bibnamefont{Stone}},
  \bibinfo{journal}{\MolP} \textbf{\bibinfo{volume}{83}}, \bibinfo{pages}{293}
  (\bibinfo{year}{1994}).

\bibitem[{\citenamefont{Lillestolen and Wheatley}(2007)}]{LillestolenW07}
\bibinfo{author}{\bibfnamefont{T.~C.} \bibnamefont{Lillestolen}}
  \bibnamefont{and} \bibinfo{author}{\bibfnamefont{R.~J.}
  \bibnamefont{Wheatley}}, \bibinfo{journal}{\JPCA}
  \textbf{\bibinfo{volume}{111}}, \bibinfo{pages}{11141}
  (\bibinfo{year}{2007}).

\bibitem[{\citenamefont{Sadlej}(1988)}]{Sadlej88}
\bibinfo{author}{\bibfnamefont{A.~J.} \bibnamefont{Sadlej}},
  \bibinfo{journal}{Coll. Czech Chem. Commun.} \textbf{\bibinfo{volume}{53}},
  \bibinfo{pages}{1995} (\bibinfo{year}{1988}).

\bibitem[{\citenamefont{Bylaska et~al.}(2006)\citenamefont{Bylaska, de~Jong,
  Kowalski, Straatsma, Valiev, Wang, Apra, Windus, Hirata, Hackler
  et~al.}}]{NWChem}
\bibinfo{author}{\bibfnamefont{E.~J.} \bibnamefont{Bylaska}},
  \bibinfo{author}{\bibfnamefont{W.~A.} \bibnamefont{de~Jong}},
  \bibinfo{author}{\bibfnamefont{K.}~\bibnamefont{Kowalski}},
  \bibinfo{author}{\bibfnamefont{T.~P.} \bibnamefont{Straatsma}},
  \bibinfo{author}{\bibfnamefont{M.}~\bibnamefont{Valiev}},
  \bibinfo{author}{\bibfnamefont{D.}~\bibnamefont{Wang}},
  \bibinfo{author}{\bibfnamefont{E.}~\bibnamefont{Apra}},
  \bibinfo{author}{\bibfnamefont{T.~L.} \bibnamefont{Windus}},
  \bibinfo{author}{\bibfnamefont{S.}~\bibnamefont{Hirata}},
  \bibinfo{author}{\bibfnamefont{M.~T.} \bibnamefont{Hackler}},
  \bibnamefont{et~al.}, \emph{\bibinfo{title}{{\sc NWChem}, a computational
  chemistry package for parallel computers, version 5.0}},
  \bibinfo{howpublished}{Pacific Northwest National Laboratory, Richland,
  Washington 99352-0999, USA.} (\bibinfo{year}{2006}).

\bibitem[{\citenamefont{Misquitta and Stone}(2013)}]{CamCASP}
\bibinfo{author}{\bibfnamefont{A.~J.} \bibnamefont{Misquitta}}
  \bibnamefont{and} \bibinfo{author}{\bibfnamefont{A.~J.} \bibnamefont{Stone}},
  \emph{\bibinfo{title}{{\sc CamCASP}: a program for studying intermolecular
  interactions and for the calculation of molecular properties in distributed
  form}}, \bibinfo{howpublished}{University of Cambridge}
  (\bibinfo{year}{2013}),
  \bibinfo{note}{http://www-stone.ch.cam.ac.uk/programs.html\#CamCASP.
  Accessed: Oct 2013}.

\bibitem[{\citenamefont{Champagne et~al.}(1998)\citenamefont{Champagne,
  Perpete, van Gisbergen, Baerends, Snijders, Soubra-Ghaoui, Robins, and
  Kirtman}}]{ChampagnePvGBSS-GRK98}
\bibinfo{author}{\bibfnamefont{B.}~\bibnamefont{Champagne}},
  \bibinfo{author}{\bibfnamefont{E.~A.} \bibnamefont{Perpete}},
  \bibinfo{author}{\bibfnamefont{S.~J.~A.} \bibnamefont{van Gisbergen}},
  \bibinfo{author}{\bibfnamefont{E.-J.} \bibnamefont{Baerends}},
  \bibinfo{author}{\bibfnamefont{J.~G.} \bibnamefont{Snijders}},
  \bibinfo{author}{\bibfnamefont{C.}~\bibnamefont{Soubra-Ghaoui}},
  \bibinfo{author}{\bibfnamefont{K.~A.} \bibnamefont{Robins}},
  \bibnamefont{and} \bibinfo{author}{\bibfnamefont{B.}~\bibnamefont{Kirtman}},
  \bibinfo{journal}{\JCP} \textbf{\bibinfo{volume}{109}}, \bibinfo{pages}{10489
  } (\bibinfo{year}{1998}).

\bibitem[{\citenamefont{Champagne
  et~al.}(1995{\natexlab{a}})\citenamefont{Champagne, Mosley, Vracko, and
  Andre}}]{ChampagneMVA95a}
\bibinfo{author}{\bibfnamefont{B.}~\bibnamefont{Champagne}},
  \bibinfo{author}{\bibfnamefont{D.~H.} \bibnamefont{Mosley}},
  \bibinfo{author}{\bibfnamefont{M.}~\bibnamefont{Vracko}}, \bibnamefont{and}
  \bibinfo{author}{\bibfnamefont{J.-M.} \bibnamefont{Andre}},
  \bibinfo{journal}{\PRA} \textbf{\bibinfo{volume}{52}}, \bibinfo{pages}{178}
  (\bibinfo{year}{1995}{\natexlab{a}}).

\bibitem[{\citenamefont{Champagne
  et~al.}(1995{\natexlab{b}})\citenamefont{Champagne, Mosley, Vracko, and
  Andre}}]{ChampagneMVA95b}
\bibinfo{author}{\bibfnamefont{B.}~\bibnamefont{Champagne}},
  \bibinfo{author}{\bibfnamefont{D.~H.} \bibnamefont{Mosley}},
  \bibinfo{author}{\bibfnamefont{M.}~\bibnamefont{Vracko}}, \bibnamefont{and}
  \bibinfo{author}{\bibfnamefont{J.-M.} \bibnamefont{Andre}},
  \bibinfo{journal}{\PRA} \textbf{\bibinfo{volume}{52}}, \bibinfo{pages}{1039}
  (\bibinfo{year}{1995}{\natexlab{b}}).

\bibitem[{\citenamefont{Saunders et~al.}(1994)\citenamefont{Saunders,
  Freyria-Fava, Dovesi, and Roetti}}]{Saunders94}
\bibinfo{author}{\bibfnamefont{V.~R.} \bibnamefont{Saunders}},
  \bibinfo{author}{\bibfnamefont{C.}~\bibnamefont{Freyria-Fava}},
  \bibinfo{author}{\bibfnamefont{R.}~\bibnamefont{Dovesi}}, \bibnamefont{and}
  \bibinfo{author}{\bibfnamefont{C.}~\bibnamefont{Roetti}},
  \bibinfo{journal}{Comp. Phys. Commun.} \textbf{\bibinfo{volume}{84}},
  \bibinfo{pages}{156} (\bibinfo{year}{1994}).

\bibitem[{\citenamefont{Drummond et~al.}(2004)\citenamefont{Drummond, Towler,
  and Needs}}]{DrummondTN04}
\bibinfo{author}{\bibfnamefont{N.~D.} \bibnamefont{Drummond}},
  \bibinfo{author}{\bibfnamefont{M.~D.} \bibnamefont{Towler}},
  \bibnamefont{and} \bibinfo{author}{\bibfnamefont{R.~J.} \bibnamefont{Needs}},
  \bibinfo{journal}{\PRB} \textbf{\bibinfo{volume}{70}},
  \bibinfo{pages}{235119(11)} (\bibinfo{year}{2004}).

\bibitem[{\citenamefont{Lopez~Rios et~al.}(2006)\citenamefont{Lopez~Rios, Ma,
  Drummond, Towler, and Needs}}]{Lopez-RiosMDTN06}
\bibinfo{author}{\bibfnamefont{P.}~\bibnamefont{Lopez~Rios}},
  \bibinfo{author}{\bibfnamefont{A.}~\bibnamefont{Ma}},
  \bibinfo{author}{\bibfnamefont{N.~D.} \bibnamefont{Drummond}},
  \bibinfo{author}{\bibfnamefont{M.~D.} \bibnamefont{Towler}},
  \bibnamefont{and} \bibinfo{author}{\bibfnamefont{R.~J.} \bibnamefont{Needs}},
  \bibinfo{journal}{Phys. Rev. E} \textbf{\bibinfo{volume}{74}},
  \bibinfo{pages}{066701} (\bibinfo{year}{2006}),
  \urlprefix\url{http://link.aps.org/doi/10.1103/PhysRevE.74.066701}.

\bibitem[{\citenamefont{Lee and Drummond}(2011)}]{LeeD11}
\bibinfo{author}{\bibfnamefont{R.~M.} \bibnamefont{Lee}} \bibnamefont{and}
  \bibinfo{author}{\bibfnamefont{N.~D.} \bibnamefont{Drummond}},
  \bibinfo{journal}{Phys. Rev. B} \textbf{\bibinfo{volume}{83}},
  \bibinfo{pages}{245114} (\bibinfo{year}{2011}),
  \urlprefix\url{http://link.aps.org/doi/10.1103/PhysRevB.83.245114}.

\bibitem[{\citenamefont{Needs et~al.}(2010)\citenamefont{Needs, Towler,
  Drummond, and Rios}}]{NeedsTDL-R10}
\bibinfo{author}{\bibfnamefont{R.~J.} \bibnamefont{Needs}},
  \bibinfo{author}{\bibfnamefont{M.~D.} \bibnamefont{Towler}},
  \bibinfo{author}{\bibfnamefont{N.~D.} \bibnamefont{Drummond}},
  \bibnamefont{and} \bibinfo{author}{\bibfnamefont{P.~L.} \bibnamefont{Rios}},
  \bibinfo{journal}{\JPhysCondMat} \textbf{\bibinfo{volume}{22}},
  \bibinfo{pages}{023201(15)} (\bibinfo{year}{2010}).

\bibitem[{\citenamefont{Trail}(2008)}]{Trail08a}
\bibinfo{author}{\bibfnamefont{J.~R.} \bibnamefont{Trail}},
  \bibinfo{journal}{Phys. Rev. E} \textbf{\bibinfo{volume}{77}},
  \bibinfo{pages}{016703} (\bibinfo{year}{2008}),
  \urlprefix\url{http://link.aps.org/doi/10.1103/PhysRevE.77.016703}.

\bibitem[{\citenamefont{Trail and Maezono}(2010)}]{TrailM10}
\bibinfo{author}{\bibfnamefont{J.~R.} \bibnamefont{Trail}} \bibnamefont{and}
  \bibinfo{author}{\bibfnamefont{R.}~\bibnamefont{Maezono}},
  \bibinfo{journal}{The Journal of Chemical Physics}
  \textbf{\bibinfo{volume}{133}}, \bibinfo{eid}{174120}
  (pages~\bibinfo{numpages}{16}) (\bibinfo{year}{2010}),
  \urlprefix\url{http://link.aip.org/link/?JCP/133/174120/1}.

\bibitem[{\citenamefont{Silberstein}(1917)}]{Silberstein17}
\bibinfo{author}{\bibfnamefont{L.}~\bibnamefont{Silberstein}},
  \bibinfo{journal}{Phil. Mag.} \textbf{\bibinfo{volume}{33}},
  \bibinfo{pages}{92} (\bibinfo{year}{1917}).

\bibitem[{\citenamefont{Applequist et~al.}(1972)\citenamefont{Applequist, Carl,
  and Fung}}]{ApplequistCF72}
\bibinfo{author}{\bibfnamefont{J.}~\bibnamefont{Applequist}},
  \bibinfo{author}{\bibfnamefont{J.~R.} \bibnamefont{Carl}}, \bibnamefont{and}
  \bibinfo{author}{\bibfnamefont{K.-K.} \bibnamefont{Fung}},
  \bibinfo{journal}{\JACS} \textbf{\bibinfo{volume}{94}}, \bibinfo{pages}{2952}
  (\bibinfo{year}{1972}).

\bibitem[{\citenamefont{Tkatchenko et~al.}(2012)\citenamefont{Tkatchenko,
  {DiStasio}, Car, and Scheffler}}]{TkatchenkoDiSCS12}
\bibinfo{author}{\bibfnamefont{A.}~\bibnamefont{Tkatchenko}},
  \bibinfo{author}{\bibfnamefont{R.~A.} \bibnamefont{{DiStasio}}},
  \bibinfo{author}{\bibfnamefont{R.}~\bibnamefont{Car}}, \bibnamefont{and}
  \bibinfo{author}{\bibfnamefont{M.}~\bibnamefont{Scheffler}},
  \bibinfo{journal}{Phys. Rev. Lett.} \textbf{\bibinfo{volume}{108}},
  \bibinfo{pages}{236402} (\bibinfo{year}{2012}),
  \urlprefix\url{http://link.aps.org/doi/10.1103/PhysRevLett.108.236402}.

\end{thebibliography}
\bibliographystyle{apsrev}

\end{document}